\documentclass[%
onecolumn,
 amsmath,amssymb,
 aps,prfluids
]{revtex4-2}

\usepackage{graphicx}
\usepackage{dcolumn}
\usepackage{bm}

\usepackage{amsmath}
\usepackage{nccmath}
\usepackage{xcolor}
\usepackage{amsmath}
\usepackage{nccmath}
\usepackage{cancel}
\usepackage{graphicx}
\usepackage{comment}
\usepackage{pgfplots}
\pgfplotsset{compat=1.18}
\usepackage{floatrow}
\usepackage{amsmath}
\usepackage{amssymb}
\usepackage{xifthen} 
\newboolean{mycondition} 
\setboolean{mycondition}{true} 
\newboolean{mycondition2} 
\setboolean{mycondition2}{false} 
\usepackage{rotating} 
\usepackage[export]{adjustbox} 
\usepackage{graphicx}
\usepackage{svg} 
\usepackage{tikz}
\usepackage{graphicx}
\usepackage{epstopdf, epsfig}

\begin{document}

\preprint{APS/123-QED}

\title{Gravity current propagating against constant and pulsating counter flows}

\author{Cem Bingol\textsuperscript{1}, Matias Duran-Matute\textsuperscript{1}, Eckart Meiburg\textsuperscript{2}, Herman J. H. Clercx\textsuperscript{1,}}
\email{Contact author: h.j.h.clercx@tue.nl}
\affiliation{\textsuperscript{1}Fluids and Flows group and J.M. Burgers Center for Fluid Dynamics, Department of Applied Physics,\\
Eindhoven University of Technology, P.O. Box 513, 5600 MB Eindhoven, The Netherlands}
\affiliation{\textsuperscript{2}Department of Mechanical Engineering, University of California at Santa Barbara, Santa Barbara, CA 93106, USA}

\begin{abstract}
This paper describes the evolution of two-dimensional (2D) gravity currents that flow against a horizontally uniform laminar pulsating flow. We study the effect of opposing mean flow amplitude and the oscillatory velocity amplitude on the evolution of the gravity current, the emergence of instabilities due to shear at the interface of heavy and light fluid and unstable density stratification near the bottom wall, and the associated density redistributions. The velocity amplitudes and the oscillation frequency are reminiscent of tidal estuarine flows. This study revealed two key processes affecting the horizontal density transport of the heavy fluid, in addition to the buoyancy-driven propagation of the gravity current. The first process concerns the presence of shear-driven Kelvin-Helmholtz (KH) billows, depending on the strength of the opposing mean flow and the thickness of the gravity current. These KH billows are generated in the inertial phase of gravity current propagation and are responsible for coherent advective transport of heavy-fluid patches away from the gravity current head. The second process is related to the lifting of the gravity current head due to differential advection near the bottom wall when the propagation direction of the gravity current and the oscillating externally imposed flow are in the same direction. It generates a layer of light fluid below the heavy fluid of the gravity current head and becomes unstable when the ambient flow opposes the gravity current propagation, generating Rayleigh-Taylor-like (RT-like) instabilities. This results in a strong vertical redistribution of light and heavy fluid. Non-hydrostatic effects, such as the presence of KH billows and RT-like instabilities, with associated vertical density transport, have significant implications for large-scale horizontal density transport and modeling of salt intrusions in rivers and estuaries.

\end{abstract}

\maketitle

\section{Introduction}

Gravity currents are ubiquitous natural flows driven by density differences in the horizontal direction. The origin of the density difference might be due to the variation in salinity, temperature, or dissolved substance in the fluid. Gravity currents are relevant for various disciplines such as environmental engineering to understand the spread of pollutants, including the prediction of oil spills that occur in the oceans, and the design of wastewater discharge systems, and coastal engineering to predict sediment transport \citep{Huppert2006}. Regardless of the cause of the density difference or the scenario in which they occur, gravity currents are driven by the same fundamental processes and exhibit similar behavior such as propagation velocity (or front velocity) and density redistributions \citep{Ungarish2020}. They have been extensively studied in various configurations to understand their dynamics, stability, and mixing properties, including non-hydrostatic effects and small-scale density redistributions \citep{Hartel2000,Necker2002,Necker2005}. Subsequent studies extended to more complex scenarios, such as a propagating gravity current over a sloping bottom \citep{Blanchette2006,Birman2007a,Dai2020,Dai2021,Maggi2023b,Martin2019,Martin2020,Zemach2019}, through an array of obstacles or over roughness elements \citep{Tokyay2011b,Maggi2022,Maggi2025b}, or into a stratified ambient fluid \citep{Maxworthy2002a,Kokkinos2023_JHR,Zahtila2024_PoF}.

Gravity currents in natural environments are often exposed to external forces, enriching the complexities of their dynamics. In coastal and estuarine regions, these forces include river discharge, tidal forcing, and wind-driven surface waves. The interplay between gravity currents and these forces has wide-ranging implications for the transport of contaminants and salt \citep{Palomar2010,Stancanelli2017ACoast,Stancanelli2018}. Recent studies investigated the effect of some of these external forces on gravity currents from a fundamental perspective. \citet{Stancanelli2018} examined the impact of short-period surface waves on gravity currents. More recently, \citet{Bingol2025} explored density redistributions and propagation of gravity currents subjected to long-period oscillatory forcing, characteristic of tidal motions.

Salt wedges in estuaries provide a particularly compelling setting, because here several external forces influence their dynamics and the mixing between salt and fresh water. In addition, salt intrusion into estuaries and rivers is a growing concern, intensified by human interventions such as dredging and sea-level rise driven by climate change. This poses a direct threat to the availability of freshwater in coastal regions. The salt wedge is particularly sensitive to two key external forces: river discharge and tidal oscillations. The river discharge generates an opposing mean flow that can arrest the salt wedge, while the tidal force causes an oscillatory flow that increases (vertical) density redistributions \citep{Geyer2011,Ralston2017} and promotes the lifting of the gravity current \citep{Bingol2025}. The combination of mean and oscillatory ambient flow, resulting in a pulsating flow, might lead to enhanced density redistributions at the interface between heavy and light fluids and in the frontal region, induce upstream density transport, and alter the propagation velocity of the gravity current. 

To address several of these processes, we performed direct numerical simulations (DNS) of the lock-exchange setup, subject to pulsating ambient flow. We varied the strengths of the mean and oscillating flow components to understand their effects on gravity current dynamics. Such a regime study using three-dimensional (3D) DNS is computationally extremely expensive due to the large difference in the time scales related to the gravity current and the ambient flow, and for that reason, we restrict ourselves at this stage to two-dimensional (2D) simulations. It serves as a starting point to explore the parameter space and to identify optimal parameter settings for a more in-depth 3D DNS campaign in the future. The opposing mean flow has a characteristic velocity $U_m$, while the oscillating velocity field corresponds to the analytical solution of Stokes' second problem and is characterized by the velocity on the free surface $U_o$. We use the dimensionless form of these characteristic velocities, denoted by the Froude numbers $Fr_m$ and $Fr_o$, respectively.

This study is divided into two parts. In the first part, our objective is to investigate the effect of only an opposing mean flow with different $Fr_m$ on the generation of instabilities at the interface between a denser and lighter fluid, the propagation of the gravity current and the transitions between different propagation phases. When $Fr_m$ increases, the front velocity of the gravity current decreases. The values of $Fr_m$ were selected such that the gravity current propagates or is marginally arrested due to the opposing flow. Although the current is expected to be slower with increasing $Fr_m$, the lighter fluid moving in the opposite direction above the density interface is expected to be faster. 
This raises questions about how the opposing flow influences the generation of instabilities at the density interface and how the density is redistributed for increasing $Fr_m$.

In the second part, our objective is to understand the influence of pulsating flow on the propagation of gravity currents and associated density redistributions, focusing on the interaction between the mean (quantified by $Fr_m$) and oscillating flow components (characterized by $Fr_o$). This investigation builds on the findings of our earlier study \citep{Bingol2025}, which explored gravity currents influenced solely by oscillatory forcing, and the present study allows us to reveal the additional effects introduced by an ambient pulsating flow. We perform simulations with varying $Fr_m$ and $Fr_o$ to evaluate the contribution of each flow component to the redistribution of the density within the gravity current, the (horizontal) transport of the density to its tail, the propagation velocity of the gravity current and the lifting of the current \citep{Bingol2025}. This will be investigated by quantifying the front position, the front velocity, the center of mass of the dense fluid in the along-channel direction with respect to the front position, and the mass flow.

\begin{figure}[!ht]
     \centering
    \caption{(a) Side view of an estuarine flow, where denser saltwater intrudes beneath lighter freshwater, forming a characteristic salt wedge. (b) Schematic of the gravity current resulting from the lock-exchange setup used in this study, where heavy fluid ($\rho_1$) and light fluid ($\rho_0$) are initially separated by a gate. The configuration assumes a simplified geometry with a horizontal, flat bottom to isolate the essential physics. Upon removal of the gate, the density difference drives a gravity current, which resembles the salt wedge dynamics illustrated in panel (a). The oscillatory velocity profile corresponding to the tidal phases with maximum velocity in both directions ($\phi = 90^{\circ}$ and $\phi = 270^{\circ}$) are shown in blue, while the laminar mean velocity profile is shown in red. The expressions for the velocity at the surface for the oscillating and mean flows are indicated above the corresponding profiles.
    }
\label{fig:estuary_flow}
\begin{tikzpicture}
\node at (0,0) {\includegraphics[width=150mm]{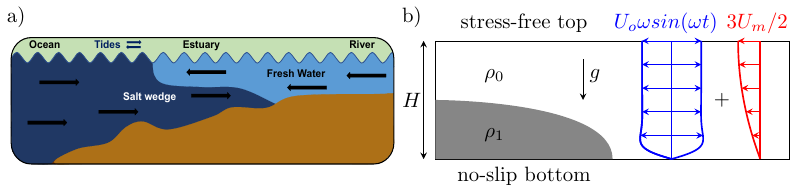}};
\end{tikzpicture}
\end{figure}

Although our study is motivated by the need to understand several aspects of the dynamics of salt intrusions in estuaries and river mouths under tidal forcing and river discharge (figure \ref{fig:estuary_flow}(a)), practical limitations force us at this stage to introduce several simplifications in the flow setup. First, we consider a flat horizontal bottom and a constant height of the fluid layer (figure \ref{fig:estuary_flow}(b)). Second, we assume oscillatory forcing (mimicking the tidal forcing) with one frequency. This is similar to what occurs in studies that only consider the dominant tidal constituent in that region. Finally, the forcing by the ambient pulsating flow field is implemented with a body force based on the associated uniform ambient pressure gradient. The impact of tides on salt intrusions is slightly different and is related to elevations of the water level and associated pressure distribution that slowly propagate upstream. This approach is not possible when the thickness of the fluid layer remains constant. However, the oscillation period of tides is much larger than the time scale associated with the propagation of the gravity current, and we assume, inspired by, for example, \citet{Gayen2010LargeCurrent}, that a time-dependent uniform (in space) pressure gradient can be applied. Although these simplifications might be limiting, they allow us to make use of a well-defined canonical flow configuration, which is advantageous for studying many aspects of gravity currents, such as propagation and mixing properties under pulsating ambient flow conditions.

Before we continue, it is worth delineating the novel aspects of the current study compared to the investigation previously reported by \citet{Bingol2025}. In that study, the propagation of gravity currents under the influence of a purely oscillating ambient flow was studied. The main emphasis was on the impact of the frequency and amplitude of the ambient flow on the dynamics of the evolving gravity current, including the redistribution of the density. That study revealed different regimes of gravity current dynamics and mass transport, including one applicable for salt intrusions in estuaries under tidal forcing. The current study is an extension of the exploration by \citet{Bingol2025} and addresses the role of river discharge on the dynamics of gravity currents, and the emergence of Kelvin-Helmholtz instabilities at the interface of heavy and light fluid, by introducing a steady mean ambient flow opposing the propagation of the gravity current. In addition, the impact of a combination of river discharge and tidal forcing, resulting in pulsating ambient flow, on the dynamics of the gravity current is explored. The frequency of the oscillating component of the ambient flow is fixed (in the regime representative of salt intrusions in estuaries), but the amplitude of the mean opposing flow (the discharge component) and of the oscillating flow (the tidal component) is varied in this study. These parameter settings result in the emergence of Rayleigh-Taylor-like instabilities on top of Kelvin-Helmholtz instabilities due to the presence of unstable density stratification near the bottom wall beneath the gravity current head.

The remainder of this paper is organized as follows. Section \ref{sec:theory} introduces the nondimensional equations of motion and dimensionless numbers governing the flow, along with a description of the numerical model, boundary and initial conditions. Section \ref{sec:results} explores the effect of the magnitude of the ambient mean flow on the propagating gravity current. In Section \ref{sec:pulsating}, the behavior of gravity currents under pulsating-flow conditions is discussed. Finally, Section \ref{sec:summary} summarizes our findings and discusses our main conclusions.

\section{Mathematical formulation and numerical setup}\label{sec:theory}

We consider a gravity current evolving in a pulsating ambient flow which is driven by an oscillating and a constant (positive) horizontal pressure gradient force,
\begin{equation}
    -\frac{1}{\rho_0}\frac{dp_0}{dx} = U_o\omega\cos(\omega t) -\frac{3\nu}{H^2}U_m~.\label{pressure}
\end{equation}

\noindent Here, $U_o$ is the velocity amplitude on the stress-free surface due to the externally imposed oscillatory flow, $\omega=2\pi/T_{osc}$ is the oscillation frequency (and $T_{osc}$ is the oscillation period), $\nu$ is the kinematic viscosity of the fluid, $H$ the depth of the fluid layer, and $U_m$ is the mean (or bulk) velocity of the laminar mean opposing flow. The mean opposing flow has a parabolic (Poiseuille) profile, which satisfies the no-slip and stress-free conditions at the bottom and top boundaries, respectively. Enforcing the no-slip boundary condition near the bottom wall for the oscillating component of the ambient flow results in a relatively thin Stokes boundary layer above the bottom wall, see Section \ref{S-2-1} for more details. The pressure gradient shown in Eq. (\ref{pressure}) results in the following pulsating free-stream velocity: $U(t)=U_o\sin(\omega t) - \frac{3}{2}U_m$. 

The motion of the gravity current can be described with the continuity, Navier-Stokes, and density transport equations. The present study considers gravity currents caused by density differences of $\mathcal{O}(1\%)$. Therefore, we can apply the Boussinesq approximation to the governing equations. These equations are made dimensionless by making the length $x_i$, velocity $u_i$ [with subscript $i$ indicating the streamwise ($x$) and wall normal ($y$) directions], time $t$, density ${\rho}$, and pressure $p$ dimensionless, as follows:

\begin{equation}
{x_i}^* = \frac{{x_i}}{{H} },
\qquad
{u_i}^* = \frac{{u_i}}{{{U}_b} },
\qquad
t^* = \frac{{t}}{{H/{U}_b} },
\qquad
    \rho ^* = \frac{ {\rho} - {\rho }_0 } {{\rho }_1 - {{\rho }}_0 },
\qquad
    p^* = \frac{{p}}{{\rho }_0 \, {{U}_b^2} },
\label{exv:eqn:UmrechnungEingangsgroesse3}
\end{equation}

\noindent where the asterisk denotes dimensionless quantities. The channel height $H$ is the characteristic length scale, the buoyancy velocity $U_b=\sqrt{{g}  ' {H}}$ is the characteristic velocity scale, ${g} \, ' = {g} \, ({{\rho }}_1-{{\rho }}_0) / {{\rho }}_0$ is the reduced gravitational acceleration (and $g$ the gravitational acceleration), and ${{\rho }}_1$ and ${{\rho }}_0$ are the densities of the heavier and lighter fluid, respectively. From this point forward, we will omit the asterisk for simplicity. The dimensionless version of Eq. (\ref{pressure}) is
\begin{equation}
    -\frac{dp_0}{dx} = \frac{Fr_o}{KC_b}2\pi\cos\left(\frac{2\pi t}{KC_b}\right) - \frac{3Fr_m}{Re}~,\label{pressure-dim}
\end{equation}
and the dimensionless form of the Boussinesq equations (with external forcing terms) is
\begin{equation}
\label{equation:dimensionless_continuity}
    \frac{\partial {u}_j }{\partial {x}_j} = 0~,
\end{equation}

\begin{equation}
\label{equation:dimensionless_ns}
    \frac{\partial  {u}_i }{\partial  t} + \frac{\partial ( {u}_i \, {u}_j)}{\partial  x_j} = - \frac{\partial  {p}}{\partial  {x}_i} +  \frac{1}{Re} \frac{\partial^2 {u}_i }{\partial {x}_j \partial {x}_j} - {\rho} \, \delta_{i2} + \left[\frac{Fr_o}{KC_b}2\pi\cos\left(\frac{2\pi t}{KC_b}\right) - \frac{3Fr_m}{Re}\right]\delta_{i1}~,
\end{equation}

\begin{equation}
\label{equation:dimensionless_transport}
    \frac{ \partial  {\rho}}{\partial  t} + \frac{ \partial ( {\rho} \, {u}_j)}{\partial  {x}_j} =  \frac{1}{Re \, Sc} \frac{ \partial ^2 {\rho}} {\partial  {x}_j \partial  {x}_j}~,
\end{equation}

\noindent where Einstein's summation convention is used. Gravity acts in the negative $y$ direction and forcing in the horizontal ($x$) direction ($\delta_{i1}$ and $\delta_{i2}$ are the Kronecker delta symbols). Here, $\rho$ is the density perturbation with respect to $\rho_0$. The dimensional quantities that define the flow are collected in Table \ref{table:dimensionfulquantities}.

\begin{table}[!ht]
\caption{\label{table:dimensionfulquantities}
Dimensional quantities that govern the fluid flow for our setup.}
\begin{ruledtabular}
\begin{tabular}{lcc}
Simulation & Description & Unit \\
\hline
$\Delta{\rho} = \rho_1-\rho_0 > 0$  & Density difference & ${\rm{kg/m^3}}$ \\
$\nu$  & Kinematic viscosity & ${\rm{m^2/s}}$ \\
$\alpha$  & Molecular diffusivity of mass & ${\rm{m^2/s}}$ \\
$g$  & Gravitational acceleration & ${\rm{m/s^2}}$ \\
$H$  & Channel height & ${\rm{m}}$ \\
$U_{m}$  & Velocity amplitude of the opposing flow & ${\rm{m/s}}$ \\
$U_{o}$  & Velocity amplitude of oscillation & ${\rm{m/s}}$ \\
$T_{osc} = \frac{2 \pi}{\omega}$  & Period of oscillation & ${\rm{s}}$ \\
\end{tabular}
\end{ruledtabular}
\end{table}

The Reynolds and Schmidt numbers are defined as
\begin{equation}
    Re = \frac{{U}_b \, {H}} {{\nu}},
\qquad
    Sc = \frac{{\nu}} {{\alpha}}~,
\label{exv:eqn:ReSc}
\end{equation}
where $\alpha$ is the molecular diffusivity of the density field. The Reynolds number indicates the relative importance of inertial forces to viscous forces, and the Schmidt number represents the ratio of momentum to density diffusion. The external forcing terms in Eq. (\ref{equation:dimensionless_ns}) introduce three more dimensionless numbers. The oscillating component of the pulsating flow is characterized by the Froude number of oscillations and the Keulegan-Carpenter number, defined as 

\begin{equation}
    Fr_o = \frac{{U}_o } {{U}_b },
\qquad
    KC_b = \frac{ {T}_{osc} \,  {U}_b}{{H}}.
\label{exv:eqn:FRKCb}
\end{equation}

\noindent These dimensionless numbers are similar to those used in our earlier study \citep{Bingol2025}, and allow a direct comparison between the present simulations with a pulsating flow and the purely oscillatory cases of earlier work. The mean opposing flow is described using the Froude number of the mean flow,

\begin{equation}
\label{equation:froude_r}
    Fr_m = \frac{{U}_m } {{U}_b }~.
\end{equation}

\subsection{Numerical model, setup, boundary conditions and parameter space}
\label{S-2-1}

A variety of approaches are available for computational studies of gravity currents. An overview is provided in the review by \citet{Meiburg2015} and a more recent example is the study by \citet{VanReeuwijk2019JFM} using SPARKLE \citep{Craske2015JFM}. We perform our 2D DNS with PARTIES \citep{Biegert2017a,Biegert2017b}, which is based on a finite-difference approach, employing the fractional step method by \citet{Moin1985}, and utilizes a third-order explicit Runge-Kutta scheme with three substeps \citep{Harten1997} to discretize the equations in time. The projection method \citep{Chorin1968NumericalEquations} is used to ensure incompressibility, and the resulting Poisson equation is treated using a fast Fourier transform (FFT) solver in each Runge-Kutta substep \citep{Biegert2017a}. The convective term in the momentum and transport equations is explicitly solved using a second-order upwind scheme. The diffusion term in the momentum and density transport equations is treated implicitly using second-order central differences in combination with a conjugate-gradient solver \citep{Saad2003IterativeSystems}. Furthermore, the code has been parallelized using the MPI library to enhance performance. The code has been extensively validated through test cases by \citet{NasrAzadani2011} and has been used in numerous studies \citep{Zhu2021RemovalCurrent, Ouillon2019TurbidityFluid, Ouillon2021GravitySources, Kollner2020b, Bingol2025}. For a detailed description of the model and the numerical schemes, the reader is referred to the earlier work by \citet{NasrAzadani2011} and \citet{Biegert2017a, Biegert2017b}.

\begin{figure*}[ht!]
  \caption{\label{fig:the_setup_1} Schematic of the lock-exchange set-up, with heavy ($\rho=1$) and light ($\rho=0$) fluid separated by a gate. The line shows the gravity current an instant after removal of the gate.}
\begin{tikzpicture}
\node at (0,0) {\includegraphics[width=134mm]{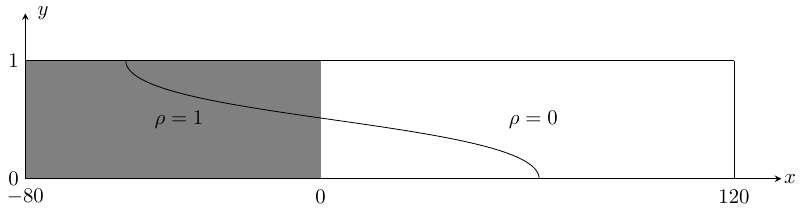}};
\end{tikzpicture}
\end{figure*}

In figure \ref{fig:the_setup_1}, we have sketched the density configuration of the lock-exchange experiment before removal of the gate. Initially, the left side of the domain ($x<0$) is occupied with heavy fluid ($\rho=1$), and the right side of the domain ($x>0$) is occupied with light fluid ($\rho=0$), separated by a gate at $x=0$. The channel height $H$ is used as the characteristic length scale, setting the domain height in dimensionless units to 1. In this dimensionless framework, the length of the heavy fluid on the left is 80 units, while the total domain length is 200 units, ensuring that the gravity current does not reach the right boundary, thereby avoiding spurious effects. On the left boundary, we have included a sponge boundary layer, which we will discuss in the next paragraph. The origin of our coordinate system is placed at the base of the gate. Positive (negative) values of $x$ indicate the right (left) side of the gate. An equidistant Cartesian mesh with a grid size $\delta_x=\delta_y=0.008$ is used for all simulations, which is sufficient for our application \citep{Bingol2025}.

A stress-free boundary condition is imposed at the rigid top, whereas a no-slip boundary condition is applied at the bottom. The density field is subjected to a no-flux condition along the bottom and top boundaries. On the left side of the domain, we use a spatial flow relaxation scheme (FRS) for the density, which is known as the sponge boundary layer \citep{Martinsen1987ImplementationModel}. The sponge boundary layer spans 10 grid cells into the domain (with grid spacing $\delta_x=0.008$, corresponding to 0.08 units). This is combined with a zero-gradient condition for the velocity. The left boundary of the domain thus represents an open boundary. This setup reduces computational costs by eliminating the need to simulate the entire length of the gravity current on the left side. On the right side of the domain, a pulsating flow profile is introduced for the velocity, while $\rho=0$. The pulsating flow profile $u_p (y,t)$, parallel to the bottom boundary, is a superposition of a steady, $u_m (y)$, and an oscillating velocity field, $u_o (y,t)$. This profile is consistent with the flow profile that would develop in a periodically forced channel flow over a no-slip bottom using the forcing term described in Eq. (\ref{equation:dimensionless_ns}). We use a laminar parabolic (Poiseuille) profile, $u_m(y)=\frac{3}{2}Fr_m y(y-2)$, and the solution of the Stokes boundary layer in a fluid layer with finite thickness (and a rigid stress-free top boundary condition) for $u_o(y,t)$, see \citet{Kaptein2019EffectRegime} for details. During the initial stages ($0\le t\le 50$) of the simulations, only the steady-flow component $u_m(y)$ is applied, allowing the gravity current to develop naturally without external pulsations influencing its early evolution. At $t=50$, the oscillatory component is superimposed on the steady ambient flow, establishing the pulsating flow conditions.

Five dimensionless numbers govern the flow: $Re$, $Sc$, $Fr_o$, $KC_b$, and $Fr_m$. To isolate the effects of the ambient mean and oscillating flow components, we systematically vary two dimensionless numbers, $Fr_m\in\{0.1,0.2,0.3,0.4\}$ and $Fr_o\in\{0.1,0.25,0.5,1\}$, while keeping $Re=3000$, $Sc=5$, and $KC_b=50$, as will be motivated below. We restricted our study to laboratory-scale 2D gravity currents with relatively small Reynolds numbers to properly resolve the smallest scales (with DNS) and to limit the necessary computational resources. This also implies that we consider only 2D (shear) instabilities and ignore 3D lobe and cleft instabilities; see \citet{Britter1978ExperimentsHead,Hallworth1996EntrainmentCurrents,Hartel2000}. We set $Re=3000$ and $Sc=5$, consistent with our earlier study on oscillatory-forced gravity currents \citep{Bingol2025}. For a more detailed discussion of the parameter space for $Re$, $Sc$, $Fr_o$, and $KC_b$, the reader is referred to Section 2.3 of \citet{Bingol2025}. They have also shown that $KC_b$ significantly influences density redistributions, revealing three distinct regimes: $KC_b \lesssim 10$, $KC_b \gtrsim 50$, and a transition regime with $10\lesssim KC_b\lesssim 50$. For $KC_b\lesssim 10$, the propagation of the gravity current shows similarities with the freely-evolving gravity current (without oscillatory forcing), but some differences are notable. The front of the gravity current exhibits instabilities, and additional redistribution of the density is observed in the presence of the externally imposed oscillating flow field. Gravity currents with $KC_b\gtrsim 50$ show density redistributions similar in shape to those observed for salt wedges in estuarine flow configurations \citep{deNijs2011}. The cases with $10\lesssim KC_b\lesssim 50$ show an intermediate behaviour, where the current front has a more inclined shape compared to the cases with $KC_b\lesssim 10$. 

The regime $KC_b\gtrsim 50$, representative of estuarine flow conditions, serves as an important benchmark for studying pulsating flow conditions. We will therefore use $KC_b=50$. We also include a set of simulations with $KC_b=0$ and $Fr_m\in\{0.1,0.2,0.3,0.4\}$, and a simulation of the freely-evolving gravity current (with $KC_b=0$ and $Fr_m=0$). These parameter settings also enable a direct comparison with the results from our earlier study of oscillatory-forced gravity currents.

\section{Gravity current with opposing mean flow}\label{sec:results}

As a first step, we performed simulations of evolving gravity currents under the influence of an opposing mean flow with $Fr_m\in\{0, 0.1, 0.2, 0.3, 0.4\}$. The selection of $Fr_m$ is made such that the gravity current is either in the arrested regime (with $Fr_m=0.4$) or propagating toward the light fluid (for $Fr_m \le 0.3$). This allows us to focus solely on how the mean flow affects key aspects of gravity current dynamics, such as the presence of shear-driven Kelvin-Helmholtz (KH) instabilities at the interface between heavy and light fluid, density redistributions around this interface, along-channel density transport within the gravity current, and its propagation characteristics. For all simulations: $Re=3000$ and $Sc=5$, facilitating a direct comparison with previous studies that focused on gravity currents under oscillatory forcing \citep{Bingol2025}. 

\subsection{General observations}\label{Sect-3Gen}
From previous studies of freely-evolving gravity currents four different propagation phases have been identified \citep{Cantero2007OnCurrents}: a brief acceleration phase immediately after gate removal, subsequently a slumping phase with constant front velocity, followed by an inertial phase and, finally, a viscous phase. As we will discuss in more detail in Section \ref{Sec3.2}, these four propagation phases are also observed for evolving gravity currents with opposing mean flow. To obtain the first preliminary insight into the evolution of gravity currents under the influence of opposing mean flow, we have added five supplementary movies, one for each $Fr_m$. For all five cases (movies 1 to 5 for $Fr_m=0$ to $0.4$, respectively), we observe the formation of KH billows during the acceleration phase after the gate is removed ($0<t\lesssim 5$). During the slumping phase, we observe weaker KH activity for $Fr_m\le 0.1$ (movies 1 and 2 for $5\lesssim t\lesssim 20$), weak and disappearing KH activity for $Fr_m=0.2$ (movie 3 for $5\lesssim t\lesssim 30$), and no significant generation of KH billows is present for $Fr_m\ge 0.3$ (movies 4 and 5 for $t\gtrsim 5$). Approximately at the transition between the slumping and inertial phase of gravity current propagation, the regular production of strong KH billows is observed for cases with $Fr_m\le 0.2$, indicating significant KH activity and the formation of a sequence of KH billows trailing behind the gravity current head. These KH billows move coherently but slowly in the tailward direction of the gravity current, taking the high-density fluid with them. These KH billows also promote strong (vertical) density redistributions. This starts at $t\approx 20$ for $Fr_m=0$ and $0.1$, and at $t\approx 30$ for $Fr_m=0.2$. This phase of KH-billow production ends at $t\approx 265$, $125$ and $85$ for $Fr_m=0$, $0.1$ and $0.2$, respectively, and approximately coincides with the transition between the inertial and viscous phase; see Section \ref{Sec3.2}.

\subsection{Density redistribution at the interface between heavy and light fluid}\label{Sect-3.1}

\begin{figure*}[ht!]
  \caption{\label{fig:Figure 3} Dimensionless density fields for different velocity amplitudes of the imposed ambient flow at $t=50$ ($L_{AR}\approx 1.3$): (a) $Fr_m=0$, (b) $Fr_m=0.1$, (c) $Fr_m=0.2$, (d) $Fr_m=0.3$, and (e) $Fr_m=0.4$. Panel (f) illustrates a comparison of the current height $h(\xi,t=50)$ for these cases. The red lines and arrows in (a)-(e) represent the horizontal velocity $u(y,t=50)$ at $x=X_{fr}-0.5$, with $X_{fr}$ the front position, and the horizontal velocity $u_m(y)$ far upstream. The value of the density (with $0 \le \rho \le 1$) is indicated by the color bar.}
\begin{tikzpicture}
\node at (0,0) {\includegraphics[width=172mm]{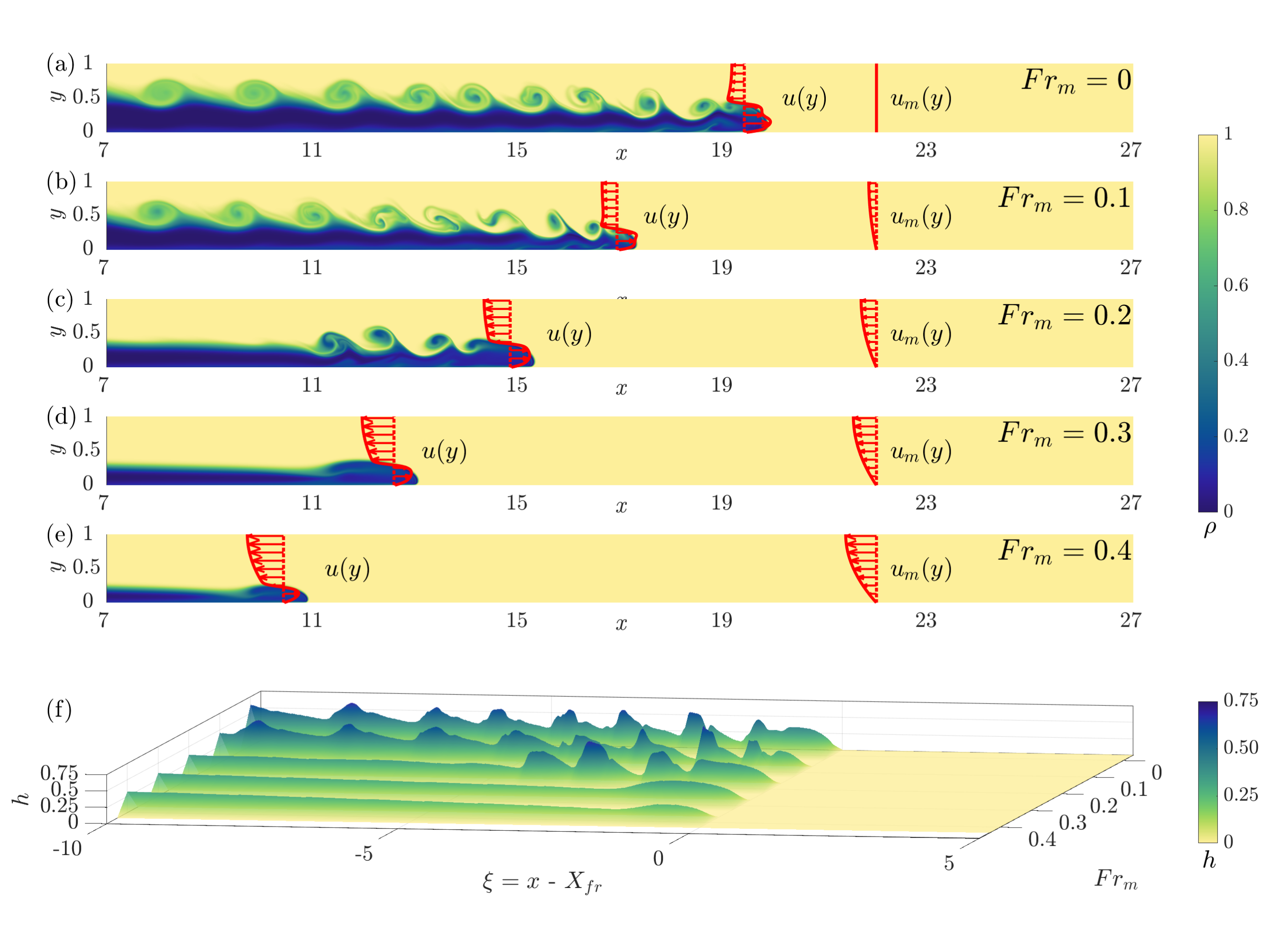}};
\end{tikzpicture}
\end{figure*}

\begin{figure}[ht!]
  \caption{\label{fig:Figure 4} The spatial distribution of $N^2$, in panels (a-e), and $S^2$, in panels (f-j) at $t=50$ for $Fr_m=0,$ 0.1, 0.2, 0.3, and 0.4, respectively ($L_{AR}\approx 2.2$). The solid red curves represent the horizontal velocity $u(y)$ at the head of the gravity current at $x=X_{fr}-0.5$ (indicated by the red vertical dotted lines). The value of the squared buoyancy frequency (with $-20 \le N^2 \le 20$) and squared shear (with $0 \le S^2 \le 200$) is indicated by the color bar.}
  \hspace*{-3mm}  
  \includegraphics[width=172mm]{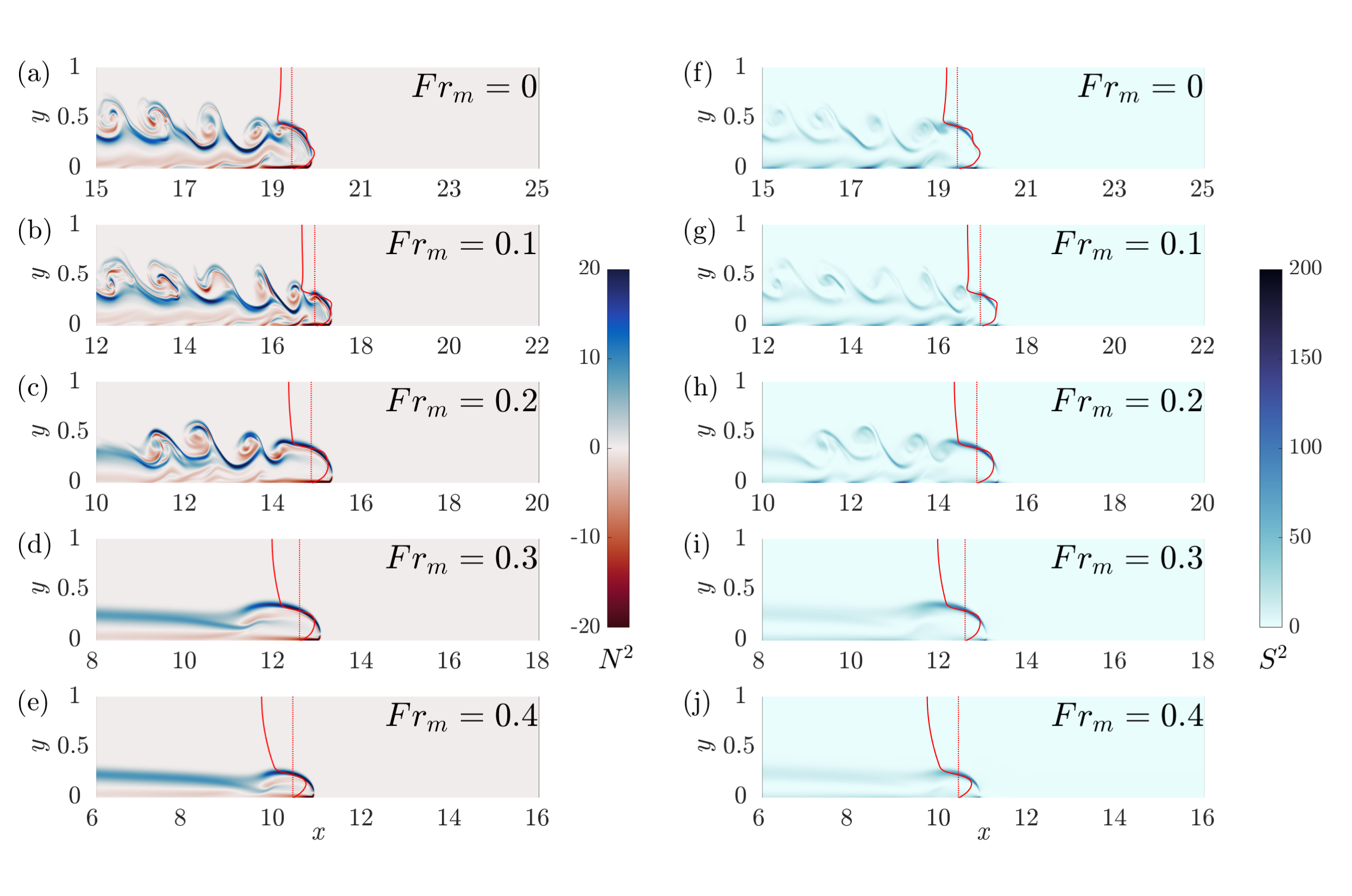}
\end{figure}

\begin{figure}[p] 
    \vspace*{-3.0cm}
  \centering
  \begin{turn}{-90}
    \begin{minipage}{\textheight} 
      \centering
      \includegraphics[trim=0cm 0cm 2.0cm 0cm, clip, width=172mm]{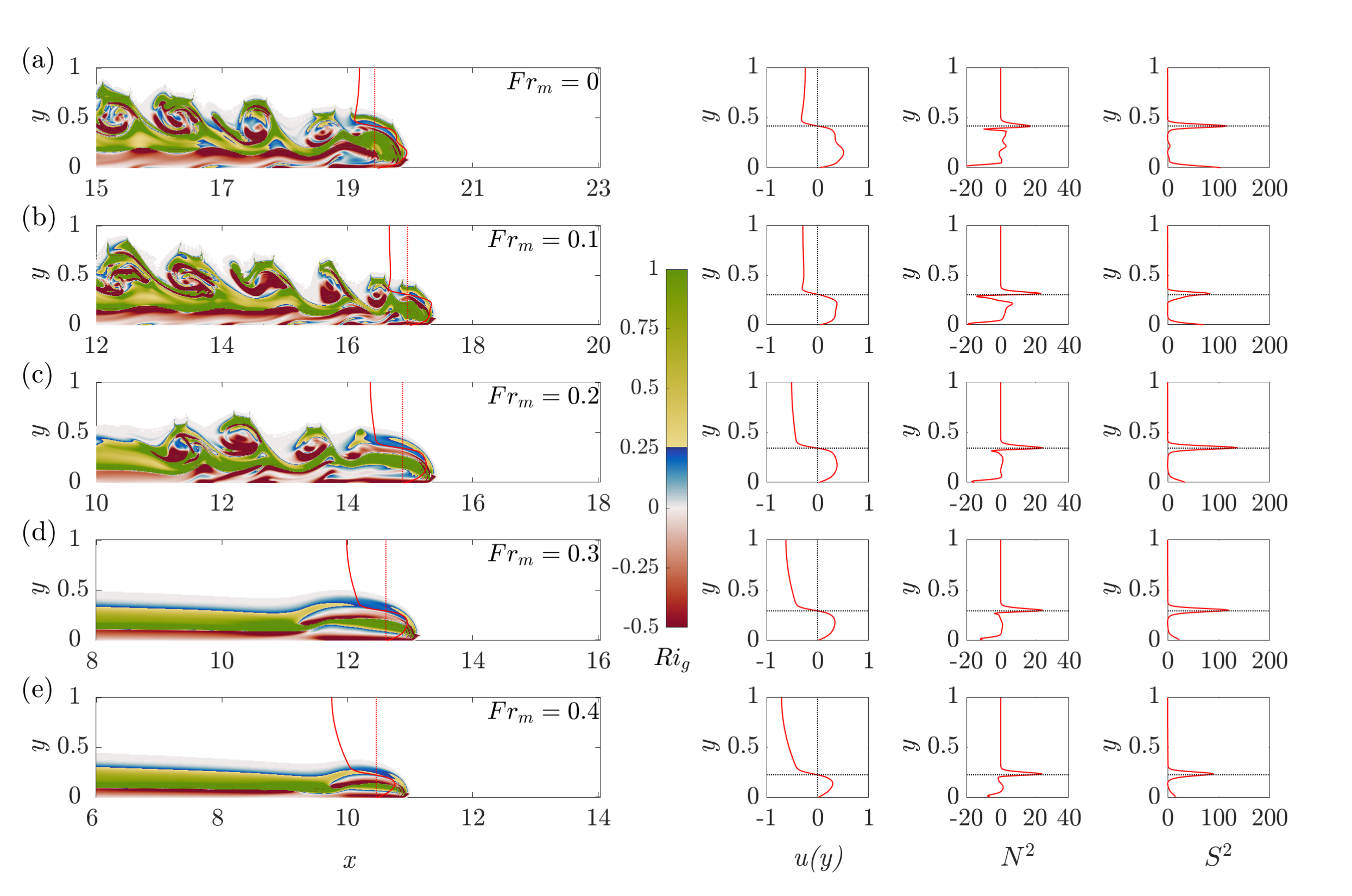}
    \hspace{-30mm}
      \caption{\label{fig:Figure 5} Gradient Richardson number, $Ri_g$, presented for $Fr_m=0$ panel (a), $Fr_m=0.1$ (b), $Fr_m=0.2$ (c), $Fr_m=0.3$ (d), and $Fr_m=0.4$ (e) at $t=50$ ($L_{AR}\approx 1.6$). The red solid curves represent the horizontal velocity $u(y)$ at the head of the gravity current for different panels ($x=X_{fr}-0.5$; indicated by the vertical red dotted line). The horizontal velocity $u(y)$, squared buoyancy frequency $N^2$ and squared shear $S^2$, 
  are plotted on the right side of each plot at $x=X_{fr}-0.5$. 
  The horizontal black dotted lines in the panels for $u(y)$, $N^2$ and $S^2$ 
  indicate the position where $u(y)=0$. The value of the gradient Richardson number (with $-0.5 \le Ri_g \le 1.0$) is indicated by the color bar.}
    \end{minipage}
  \end{turn}
\end{figure}

\begin{figure}[ht!]
  \caption{Dimensionless density fields for different $Fr_m$ ($L_{AR}\approx 5.6$). Left: $t<t_{tr}$ with formation of KH billows. Right: $t>t_{tr}$ where the KH instability is suppressed. Panels (a-b): $Fr_m=0$ and $t_{tr}\approx 265$; panels (c-d): $Fr_m=0.1$ and $t_{tr}\approx 125$; panels (e-f): $Fr_m=0.2$ and $t_{tr}\approx 85$. The value of the density (with $0 \le \rho \le 1$) is indicated by the color bar.\label{fig:Figure 6} }
  \hspace*{-3mm}  
  \includegraphics[width=160mm]{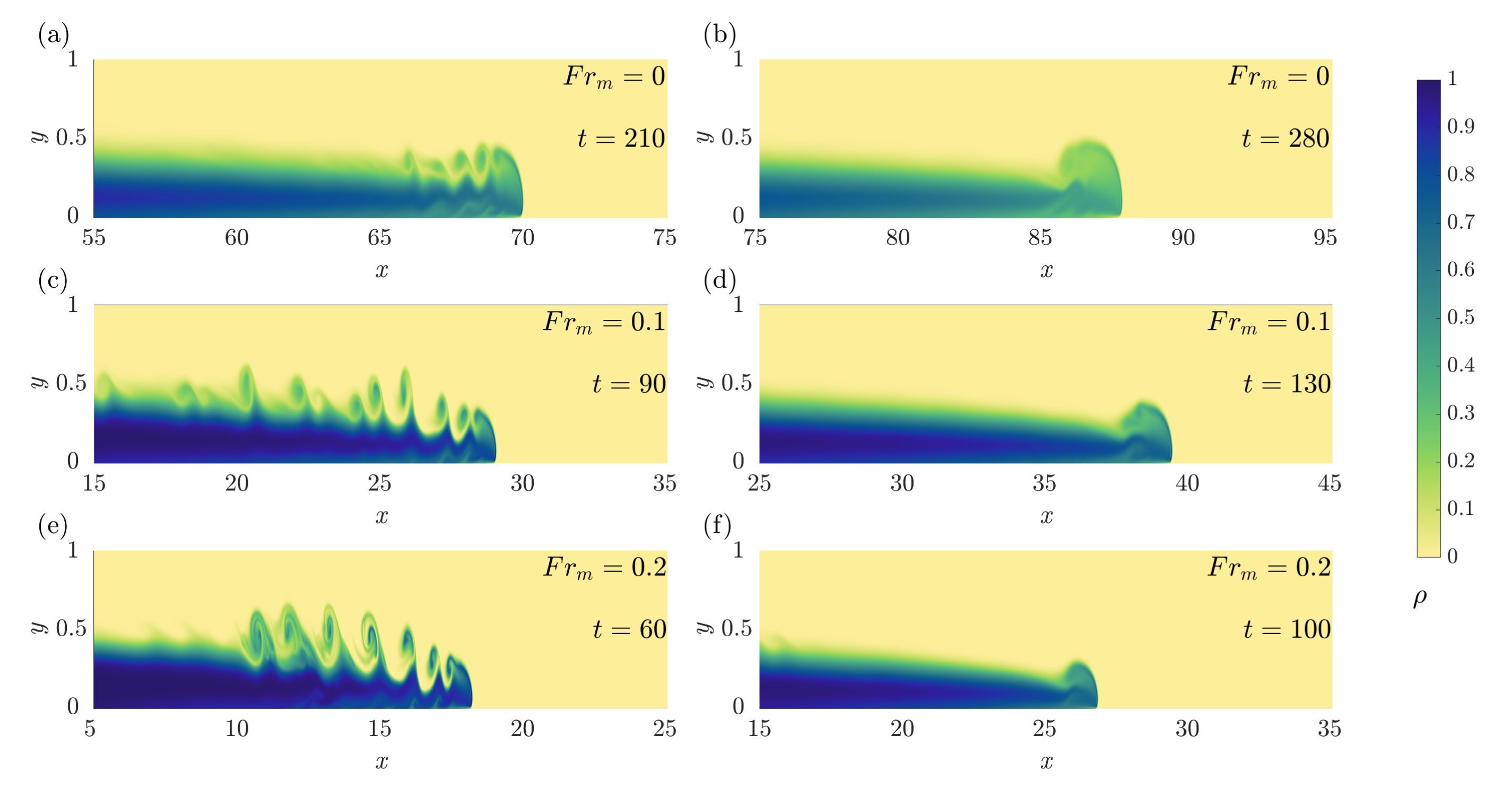}
\end{figure}

Figure \ref{fig:Figure 3} provides us with a qualitative comparison of the behavior of the gravity current for varying $Fr_m$ using snapshots of the density field at $t=50$.\footnote{The horizontal and vertical scales of the snapshots in figures \ref{fig:Figure 3}-\ref{fig:Figure 6} and \ref{fig:Figure 10} are generally not equal. For example, in figure \ref{fig:Figure 3}, the horizontal axis is compressed by a factor of $1.3$ and the aspect ratio of the vertical to horizontal unit distance is $L_{AR}\approx 1.3$.} We immediately observe that the average propagation velocity decreases with increasing $Fr_m$, which can be anticipated given the increasing strength of the opposing flow. Additionally, the emergence of KH billows is absent when $Fr_m\gtrsim 0.3$. This observation hints at a $Fr_m$-dependent balance between destabilizing velocity gradients and stabilizing density gradients, which will be further explored in the following using the gradient Richardson number. Furthermore, we observe that the current height decreases with increasing $Fr_m$ (see figure \ref{fig:Figure 3}(a)-(e)). An experimental study on gravity currents subjected to uniform opposing flow conditions has also reported a negative correlation between the velocity amplitude of the opposing flow and the current height \citep{Riddel1970}.

The current height is generally used as a measure for the vertical position of the gravity current interface between heavy and light fluids, and various definitions have been proposed \citep{Shin2004,Anjum2013,Bingol2025}. \citet{Anjum2013} introduced an expression for the local current height $h_{a}(x,t)$, defined as

\begin{equation}
\label{eq:cm}
    h_{a}(x,t) = 2h_{cm}(x,t) = 2\frac{\int_{0}^{1} \rho (x,y,t) \, y \, dy}{\int_{0}^{1} \rho (x,y,t) \, dy}~,
\end{equation}

\noindent with $h_{cm}(x,t)$ the dimensionless center of mass of the current in the vertical direction. The definition assumes that $h_{a}(x,t)$ is not evaluated beyond the front position $X_{fr}$ where $h_{a}(x,t)=0$. We determined $X_{fr}$ by approaching the gravity current from the right side and finding the horizontal position $X_{fr}(t)=x$ where $\int_0^1\rho(x,y,t)dy=0.005$ for the first time. We varied this threshold between $0.01$ and $0.0025$, affecting the results for $X_{fr}$ by less than 1\%.

Recently, \citet{Bingol2025} demonstrated that $h_a(x,t)$ overestimates the current height when there is a significant amount of lighter fluid below the gravity current, for example, due to the lifting of the gravity current under oscillatory forcing conditions (without laminar mean opposing flow, thus $Fr_m=0$). For the gravity current under pulsating flow conditions, we anticipate similar lifting to take place. Therefore, we use the measure for the local current height $h(x,t)$ proposed by \citet{Bingol2025}, which is the sum of the vertical center of mass $h_{cm}(x,t)$ and the local current height of the heavy current above $h_{cm}(x,t)$, yielding

\begin{equation}
    h (x,t) = h_{cm}(x,t)  + 2 \, \frac{\int_{h_{cm}(x,t)}^{1} \rho (x,y,t) \, (y-h_{cm}(x,t)) \, dy}{\int_{h_{cm}(x,t)}^{1} \rho (x,y,t) \, dy}~.
    \label{curr-height}
\end{equation}

\noindent This expression implies that $h(x,t)=0$ for $x \ge X_{fr}$, and when there is no light fluid below the heavy fluid, $h(x,t) = 2h_{cm}(x,t)$. To compare the current heights, we introduce a horizontal coordinate with respect to the front position, $\xi(t)=x-X_{fr}(t)$. The results shown in figure \ref{fig:Figure 3}(f) illustrate the decrease in current height with increasing $Fr_m$. 

To perform a more detailed analysis, we calculated the average height of the gravity current in the front region using the expression:

\begin{equation}
    h_{f}(t)=\frac{1}{\Delta L}\int_{X_b}^{X_{fr}} h(x,t) dx,
    \label{curr-front}
\end{equation}

\noindent with $X_b=X_{fr}-\Delta L$ and $\Delta L$ the length of the front region. We use $\Delta L=15$ as the optimal value \citep{Bingol2025}. After time averaging, we identified the following empirical relationship: $h_{f}\simeq 0.5(1-Fr_m)$, showing a negative correlation between current height and the strength of the opposing flow, consistent with previous experiments \citep{Riddel1970}. 

A further examination of the density field near the interface illustrates the role of KH billows in driving the redistribution of heavy and light fluid, observed for $Fr_m \lesssim 0.2$. These billows originate from the shear induced by the exchange flow between the dense fluid moving in the positive $x$ direction near the bottom of the channel and the light fluid moving in the opposite direction, above the gravity current. The shear promotes KH instabilities and the subsequent formation of KH billows near the gravity current head, whereas stable stratification opposes their development. The local value of shear $S(x,y,t)$ (for a 2D flow field) is quantified by 
\begin{equation}
S^2(x,y,t)= \left(\frac{\partial u(x,y,t)}{\partial y}\right)^2~,
\end{equation}
while the local value of stratification is represented by the buoyancy frequency $N(x,y,t)$, defined such that
\begin{equation}
N^2(x,y,t) = -\frac{\partial \rho(x,y,t)}{\partial y}~.
\end{equation}
The quantity $S^2(x,y,t)$ is always positive, but $N^2(x,y,t)$ can be negative (and the fluid has a locally unstable density stratification). To further examine the interaction between shear and stratification, we computed the spatial distribution of $N^2$ and $S^2$ at $t=50$ (figure \ref{fig:Figure 4}) near the head of the gravity current. 

Large positive values of $N^2$ (in blue) at the head of the gravity current and further downstream nicely indicate the position of the interface (see figure \ref{fig:Figure 4}(a)-(e)). In contrast, negative values of $N^2$ (in red) indicate locations with an unstable density distribution with a denser fluid above lighter fluid. This unstable configuration is evident in regions where KH billows are present, as well as just above the no-slip bottom wall as a result of limited entrainment of lighter fluid beneath denser fluid during passage of the gravity current head. Additionally, the horizontal velocity profile, $u(y)$ (red curve in figure \ref{fig:Figure 4}), taken slightly behind the front position at $x=X_{fr}-0.5$ reveals that the velocity gradient is highest near the interface. Consequently, $S^2$ also reaches maximum values in this region for all cases (see figure \ref{fig:Figure 4}(f)-(j)). 

In general, the bulk Richardson number and the stability Froude number
\citep{Long1956,Lawrence1991TheInterface} are widely used measures to assess the stability properties of flows. The bulk Richardson number characterizes the balance between stratification and shear (similar in form to the local gradient Richardson number, $Ri_g$, defined below in Eq. (\ref{rig_eq})), but is computed using domain-averaged values under the assumption of uniform stratification and shear. In contrast, the stability Froude number reflects the potential for interfacial instability (again conceptually related to $Ri_g$ through linear stability analysis), but it quantifies the stratification–shear balance based on the average properties of the two fluid layers (for our case, the light and dense fluid layers). The bulk Richardson number is used to parameterize shear-driven turbulence \citep{Smyth2000LengthLayers,Mashayek2017RoleLayers} while the stability Froude number is used to assess the growth of instability \citep{Atoufi2023StratifiedInstabilities} in stratified exchange flows. They are better suited for cases with limited data resolution or a relatively steady stratification-shear balance. 

In contrast, given the intrinsic unsteadiness of the flow in our case and the strong spatial dependence of stratification and shear in the propagating gravity current, a more localized measure is needed to capture the location and growth of instabilities. For this purpose we introduce the local gradient Richardson number, 
\begin{equation}
\label{rig_eq}
Ri_g(x,y,t)=\frac{N^2(x,y,t)}{S^2(x,y,t)}~,
\end{equation}
which represents a local balance between stratification and shear. According to linear stability analysis, a necessary (but not sufficient) condition for instabilities to grow and develop into KH billows is that the gradient Richardson number be less than the critical Richardson number, thus $Ri_g(x,y,t) \le Ri_c=0.25$ somewhere in the domain \citep{Miles1961OnFlows}.

The gradient Richardson number exhibits a strong spatial and temporal dependence, see figure \ref{fig:Figure 5}(a)-(e) for the spatial variability, taken at the same instant as the results shown in figures \ref{fig:Figure 3} and \ref{fig:Figure 4}. The generation of KH billows occurs near the gravity current head; thus, we will initially focus on the vertical cross section at $x=X_{fr}-0.5$, indicated by the vertical red dotted line. This choice is somewhat arbitrary, considering also the spatio-temporal behavior of $Ri_g(x,y,t)$ in mind, thus the discussion has a qualitative character. Figure \ref{fig:Figure 5} shows that near the interface (where $u(y)=0$) large positive values are observed for both $S^2$ (large shear) and $N^2$ (stable density gradient). Near $x\approx X_{fr}-0.5$ and $u(y)\approx 0$, thus the small region around the intersection of the solid red curve and the vertical red dotted line in figure \ref{fig:Figure 5} left column, we observe that $Ri_g$ changes from blue/red (thus $Ri_g\lesssim 0.25$) for $Fr_m\lesssim 0.2$ to yellowish (thus $Ri_g\approx 0.25$) for $Fr_m=0.4$. This is corroborated by the maximum values for $N^2$ (almost unchanged for $Fr_m\ge 0.2$) and $S^2$ (decreasing with increasing $Fr_m$), thus increasing $Ri_g$ with increasing $Fr_m$. This supports the presence of KH billow formation only for $Fr_m\le 0.2$.

For $Fr_m=0.2$ and $t\lesssim 85$, KH billows form near the head of the gravity current (figure \ref{fig:Figure 3}(c) and figure \ref{fig:Figure 6}(e)). As time progresses, the height and front velocity of the gravity current decrease, while the height of the upper layer increases. Consequently, the mean velocity in the upper layer decreases given that the flow rate in this layer is constant. Hence, the mean velocity in both the lower and upper layers decreases over time, leading to a reduction in shear between the two layers. This suppresses the formation of KH billows, and for the case with $Fr_m=0.2$ at $t=100$, in fact, no generation of KH billows is observed; see figure \ref{fig:Figure 6}(f). A transition occurs between the two behaviors for $t_{tr}\approx 85$. A similar evolution of the propagating gravity current is observed for $Fr_m=0$ and $0.1$, although at later times (figure \ref{fig:Figure 6}(a)-(d)). The transition then occurs at $t_{tr}\approx 125$ for $Fr_m=0.1$ and $t_{tr}\approx 265$ for $Fr_m=0$. Animations of the density field evolution for these three cases, illustrating the formation of KH billows and the cessation of it for $t\gtrsim t_{tr}$, are provided in the supplementary movies discussed in Section \ref{Sect-3Gen}. 

As a proxy for assessing vertical density transport by advection, we consider horizontally averaged density variance 
\begin{equation}\label{eq:rho-average}
\sigma_{\rho,\Delta X}^2(y,t)=\langle (\rho(x,y,t)-\langle\rho\rangle_{\Delta X}(y,t))^2\rangle_{\Delta X}~,
\end{equation} 
covering a range with size $\Delta X$ just behind the gravity current head. When KH billows are present, their advective properties cause significant non-zero $\sigma_{\rho,\Delta X}^2(y,t)$ in a large part above the local current height $h(x,t)$. Since this does not happen without KH billows, we conclude that the vertical advective transport of the density is supported by the presence of KH billows only for $t\lesssim t_{tr}$. This has immediate implications for horizontal density transport, which will be discussed in Section \ref{Sec3.2}.

\begin{figure}[ht!]
  \caption{\label{fig:Figure 7} Spatio-temporal diagram ($x$-$t$) of the gradient Richardson number, $\langle Ri_g\rangle_{if}$, for (a) $Fr_m=0$, (b) $Fr_m=0.1$, and (c) $Fr_m=0.2$ measured for $x\le X_{fr}(x,t)-0.5$. Only the critical $\langle Ri_g\rangle_{if}<0.25$ is shown (and $\langle Ri_g\rangle_{if}\le 0$ is colored yellow). The red dashed line shows $X_{fr}(x,t)$.}
  \hspace*{-3mm}  
  \includegraphics[width=172mm]{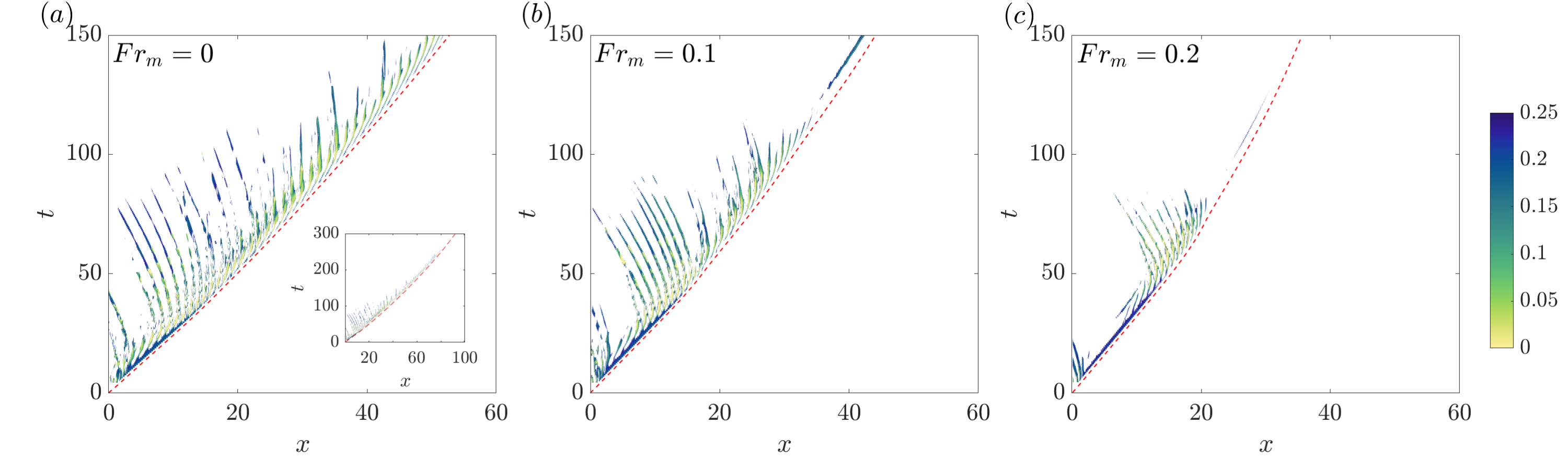}
\end{figure}

To better understand the influence of opposing flow (by applying different $Fr_m$) on the KH billow generation mechanism from a spatio-temporal viewpoint, we investigated the local gradient Richardson number averaged over the thickness of the interface, 
\begin{equation}
\langle Ri_g\rangle_{if}(x,t)=\frac{1}{h_{i_+}-h_{i_-}}\int_{h_{i_-}}^{h_{i_+}} Ri_g(x,y,t)dy~,
\end{equation}
near the gravity current head as a function of the horizontal coordinate and time. The interface thickness, $h_{i_+}-h_{i_-}$, is estimated with the width of the peaks of $S^2$ around $u(y)=0$, see figure \ref{fig:Figure 5}. The qualitative description from the previous paragraph is supported with space-time diagrams of $\langle Ri_g\rangle_{if}$ in figure \ref{fig:Figure 7}, where only values in the range $0<\langle Ri_g\rangle_{if} \le 0.25$ are shown. These diagrams reveal by the curved spikes, which serve as proxies for the presence of KH billows, that instabilities are generated approximately until $t_{tr}$ for $Fr_m=0$, $0.1$, and $0.2$. When these curved spikes disappear, like for $t\gtrsim t_{tr}\approx 85$ in figure \ref{fig:Figure 7}(c), $\langle Ri_g\rangle_{if}\gtrsim 0.25$ and no more instabilities are present, and no new KH billows are formed. These diagrams support our observations (figure \ref{fig:Figure 6}) that instabilities are generated for a shorter duration as $Fr_m$ increases. Therefore, increasing $Fr_m$ has a stabilizing effect and suppresses the generation of KH billows at the interface of the gravity current head and reduces the vertical density transport by advection. 

\subsection{Transport of high-density fluid}\label{Sec3.2}

The transport of the substance responsible for the density difference (e.g. salt, heat, sediment, or pollutants) is of interest for many applications. Therefore, we focus on understanding how the mean opposing flow (quantified by $Fr_m$) affects mass transport by gravity currents. We quantify this with the front position, which indicates the extent of the current, the front velocity, the mass flow through the gate, and the horizontal position of the center of mass of the high-density fluid.

We first examine the extent of the gravity current, represented by the front position $X_{fr}$ and the propagation velocity $U_{fr}$. For freely-evolving gravity currents ($Fr_m=0$), the propagation of the gravity current consists of four different phases: acceleration, slumping, inertial, and viscous phases \citep{Cantero2007OnCurrents}. Each of these phases has specific scaling characteristics for the front position $X_{fr}(t)$ and the front velocity $U_{fr}(t)$ as a function of time. These scaling regimes allow one to identify the phases that the gravity current is experiencing. The behavior of the front velocity $U_{fr}$ provides an alternative for $X_{fr}$ to identify the different phases. The front velocity is calculated as 


\begin{equation}
    U_{fr}(t) = \frac{\mathrm{d} X_{fr}(t)}{\mathrm{d} t}.
\end{equation}

\noindent We first consider the freely-evolving gravity current ($Fr_m=0$), see supplementary movie 1. After removal of the gate, we enter a brief acceleration phase of the gravity current and arrive quickly in the slumping phase. The slumping phase is characterized by a constant front velocity $U_{fr}$ and $X_{fr}\propto t$. In the inertial phase, the height of the current head, $h_{f}(t)$, scales with $t^{-2/3}$ \citep{Ungarish2020} while the front velocity $U_{fr}\propto t^{-1/3}$ \citep{Cantero2007OnCurrents}. After the inertial phase, the gravity current experiences the viscous phase in which buoyancy and viscous forces are in balance. \citet{Hoult1972OilSea} considered the viscous forces acting between the interface of heavy and light fluids and demonstrated that $U_{fr}\propto t^{-5/8}$. \citet{Huppert1982TheSurface} performed a revised analysis accounting for the viscous effect on a rigid horizontal surface and showed that $U_{fr} \propto t^{-4/5}$.

\begin{figure}[ht!]
  \caption{\label{fig:Figure 8} Time evolution of (a) the front position $X_{fr}$, (b) the front velocity $U_{fr}$, and (c) the Reynolds number based on the height and velocity of the gravity current head, $Re_f=U_{fr}h_f/\nu$, for $0.5 \le t \le 400$ and varying values of $Fr_m$. In the slumping phase, $X_{fr}(t)\propto t$ as shown with the black dotted line in panel (a). Additionally, we have added short dotted lines to approximately indicate the inertial phase for $Fr_m=0$, $0.1$ and $0.2$ (red dotted lines) and the viscous phase proposed by \citet{Huppert1982TheSurface} for $Fr_m=0.3$ and $0.4$ (purple dotted line). Panel (b) shows the scaling laws for the front velocity during different propagation phases: the slumping phase ($t^0$), the inertial phase ($t^{-1/3}$), the viscous phase by \citet{Hoult1972OilSea} ($t^{-5/8}$), and the one by \citet{Huppert1982TheSurface} ($t^{-4/5}$), represented by black, red, blue, and purple dotted lines, respectively. In panel (c), the horizontal blue line indicates $Re_f \approx 90$, marking the transition to the viscous phase.
  }
  \hspace*{-3mm}  
  \includegraphics[width=150mm]{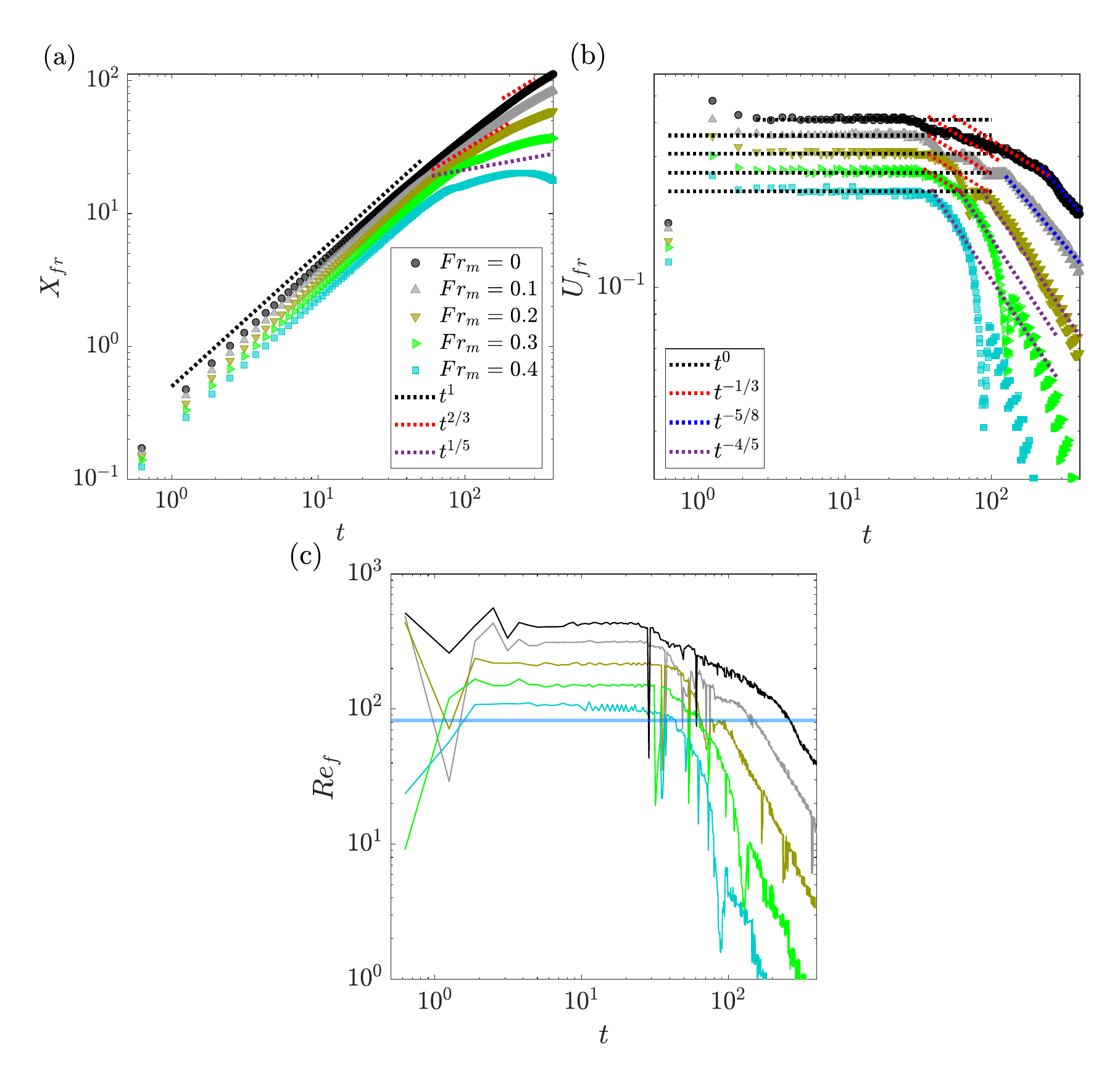}
\end{figure}

The scaling characteristics of $X_{fr}(t)$ and $U_{fr}(t)$, mentioned above for the freely-evolving gravity current, are an ideal starting point to explore the propagation phases and associated scaling properties of gravity currents affected by a mean opposing flow. Figure \ref{fig:Figure 8}(a) illustrates $X_{fr}$ for the explored range of $Fr_m$ and shows that $X_{fr}(t)\propto t$ for $t\lesssim 50$. For $Fr_m \le 0.3$, the front position advances throughout the simulation, but for $Fr_m=0.4$ the gravity current is arrested, with the front position $X_{fr}$ remaining nearly constant for $200 \le t \le 400$. In the acceleration phase, $U_{fr}$ increases rapidly and reaches a constant value (figure \ref{fig:Figure 8}(b)). During the slumping phase, the front velocity remains constant (and $X_{fr}\propto t$). In addition, the Reynolds number based on height $h_f$ and front velocity $U_{fr}$, defined as $Re_f = Re \, (h_f/H) \, (U_{fr}/U_{b}) = h_f U_{fr}/\nu$, remains constant during the slumping phase (figure \ref{fig:Figure 8}(c)), which implies that the height of the current head also remains constant. From figure \ref{fig:Figure 8}(b), we find $U_{fr} \approx 0.41(1-Fr_m^{0.9})$.

After the slumping phase, the gravity current may transition to either the inertial phase, followed by the viscous phase, or skip the inertial phase altogether and enter directly into the viscous phase. These transitions are most clearly visible for the front velocity $U_{fr}(t)$, but as indication of the presence of the inertial and viscous phases in the plots of $X_{fr}(t)$ we have also added dotted lines representative for the inertial scaling ($\propto t^{2/3}$) and the viscous scaling ($\propto t^{1/5}$) by \citet{Huppert1982TheSurface}. As shown in figure \ref{fig:Figure 8}(b), the inertial phase exists for gravity currents when $Fr_m \lesssim 0.3$. The observed behavior in the viscous phase depends on $Fr_m$. For simulations with $Fr_m=0$ and 0.1, $U_{fr}$ follows the scaling proposed by \citet{Hoult1972OilSea} ($U_{fr} \propto t^{-5/8}$). However, for $Fr_m \geq 0.2$, we obtain $U_{fr} \propto t^{-4/5}$, in line with the scaling suggested by \citet{Huppert1982TheSurface}. This suggests that for $Fr_m \le 0.1$, the viscous forces at the interface are more dominant, while for $Fr_m \ge 0.2$, the viscous forces at the bottom become more dominant. This can be attributed to the decrease in the front velocity as $Fr_m$ increases (owing to the enhanced opposing flow), which thickens the boundary layer. This makes the viscous effects at the bottom a dominant factor that affects the propagation of the gravity current, aligning with the argument of \citet{Hallez2009ACurrents} on the reason for two different scaling behaviors for the viscous phase seen by \citet{Cantero2007OnCurrents}. For $Fr_m=0.4$, the gravity current completely skips the inertial phase and transitions directly from the slumping phase to the viscous phase, following the scaling with $U_{fr} \propto  t^{-4/5}$.

The transition to the viscous phase occurs as $h_f$ and $U_{fr}$ decrease, and directly affect $Re_f$, see figure \ref{fig:Figure 8}(c). The transition to the viscous propagation phase occurs for $Re_f\approx 90$ (see figure \ref{fig:Figure 8}(b) and the horizontal blue line in \ref{fig:Figure 8}(c)). Furthermore, these data clearly illustrate that as $Fr_m$ increases, the transition to the viscous scaling regime occurs earlier. Inspection of figures \ref{fig:Figure 8}(b) and \ref{fig:Figure 8}(c) reveals that the transition to the viscous regime occurs approximately at $t_{tr}$, introduced in Section \ref{Sect-3.1}. Hence, the cessation of the generation of KH billows and the reduction of advective density transport occur when the propagation phase enters the viscous regime. The supplementary movies also show that the transition from the slumping phase to the inertial phase (typically in the range $20\lesssim t\lesssim 40$) is accompanied by an intensification of the formation of KH billows ($Fr_m\le 0.3$). 

\begin{figure}[ht!]
  \caption{\label{fig:Figure 9} Time evolution of (a) the average density $M_g(t)$ on the right side of the gate between $0\le x\le X_{fr}(t)$, and (b) the center of mass of high-density fluid on the right side of the gate normalized with the front position, $L_m(t)$, both presented for varying values of $Fr_m$.}
  \hspace*{-3mm}  
  \includegraphics[width=150mm]{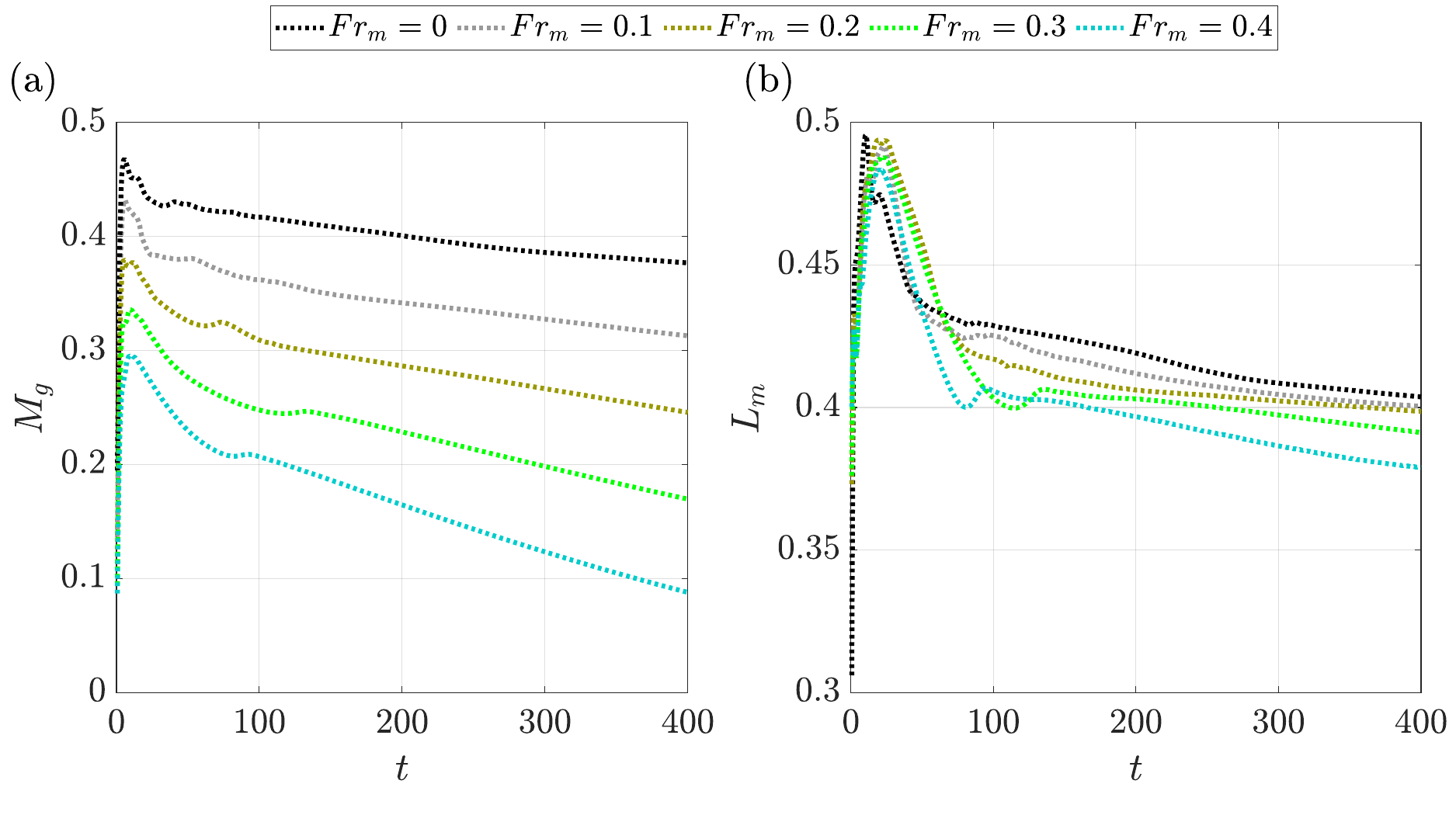}
\end{figure}

Variations in the opposing flow strength, represented by $Fr_m$, influence the transport of high-density fluid within the gravity current. The amount of high-density fluid transported through the gate represents the primary intrusion of dense fluid into the region initially occupied by lighter fluid. To gain a deeper understanding of mass transport into the critical region (the region beyond the gate, initially filled with light fluid), we examine two metrics: the average density on the right side of the gate (between $0\le x \le X_{fr}(t)$),
\begin{equation}\label{mass-lock}
    M_{g} (t) = \frac{1}{X_{fr}(t)}\int_{0}^{L}\int_0^1 \rho (x,y,t) \,dx \,dy~,~
\end{equation}
and the normalized center of mass of dense fluid on the right side of the gate,

\begin{equation}\label{equation:L_m}
    L_{m} (t) = \frac{1}{X_{fr}(t)}\frac{\int_{0}^{L}\int_0^1 \rho (x,y,t) x \,dx \,dy~}{\int_{0}^{L}\int_0^1 \rho (x,y,t) \,dx \,dy~}~, 
\end{equation}
which measures whether the dense fluid is mostly located in the gravity current head or mostly farther behind the head. As a reference, we consider the idealized case of a gravity current with uniform density $\rho=1$ and height $h(x,t)=0.5$ for $0\le x\le X_{fr}$. In that case: $M_g(t)=0.5$ and $L_m(t)=0.5$.

Figure \ref{fig:Figure 9}(a) illustrates that the total mass of dense fluid per unit length on the right side of the gate, $M_g(t)$, decreases almost linearly with increasing $Fr_m$. This suggests that $M_g(t)\approx 0$ for $Fr_m$ in the range $0.5-0.7$ and indicates that the transport of density into the critical region beyond the gate is strongly suppressed by a sufficiently strong opposing flow. Figure \ref{fig:Figure 9}(b) shows a reduction in the normalized center of mass of dense fluid, $L_m(t)$, thus dense fluid tends to propagate less far beyond the gate. A sharp decrease of $L_m(t)$ for $20\lesssim t\lesssim 100$ is observed (see figure \ref{fig:Figure 9}(b)) which largely coincides with the inertial phase. This decrease is associated with the presence of the opposing flow and, for $Fr_m\lesssim 0.2$, the generation of KH billows. The KH billows, in addition to promoting strong (vertical) density redistributions, move coherently but slowly in the tailward direction of the gravity current; see Section \ref{Sect-3Gen}. These KH billows take the high-density fluid with them. The impact of KH billows and opposing flow are difficult to disentangle, but the trend with increasing $Fr_m$ suggests dominance of density transport by KH billows ($Fr_m=0$) or by opposing flow ($Fr_m=0.4$). $Fr_m=0.2$ represents an intermediate case, with KH billows still present but relatively weak, and stronger opposing flow. The subsequent decay for $t\gtrsim 100$ (mainly in the viscous phase of propagation) is mainly due to the diffusive vertical transport of the density and the subsequent advection of the density by horizontal flow. 

\section{Evolution of gravity currents against a pulsating flow} \label{sec:pulsating}

To generate pulsating flow conditions, we introduce an oscillating flow component alongside the opposing mean flow. This adds one dimension to the parameter space, as we can vary both $Fr_o$ and $Fr_m$ (in our study $KC_b=50$ and remains constant). This section presents results from 16 simulations using $Fr_m=0.1$, $0.2$, $0.3$, $0.4$ and $Fr_o=0.1$, $0,25$, $0.5$, $1.0$. For $t \leq 50$, the externally imposed flow consists solely of the mean flow profile, $u_m(y)=\frac{3}{2}Fr_m y(y-2)$. At $t=50$, for all $Fr_m$ beyond the slumping phase of the gravity current propagation, the externally imposed mean flow is combined with the oscillating component $u_o(y,t)$, which corresponds to the solution of the Stokes boundary layer, as described by, e.g., \citet{Kaptein2019EffectRegime}. Our primary focus is on the results for $t \geq 50$. 

We will provide qualitative insights into (horizontal) density transport within gravity currents and at the interface between heavy and light fluids by examining the evolution of density fields during different phases of the oscillation cycle under pulsating-flow conditions. We compare these results with those of purely oscillating ambient flows to highlight key differences \citep{Bingol2025}. To better understand the effect of the oscillatory component of the pulsating ambient flow, we introduce the cycle-averaged front position and propagation velocity of the gravity current. In addition, we present the net mass transport of the dense fluid and discuss its effect on horizontal density transport within the gravity current. 

\subsection{Propagation of the gravity current during an oscillation cycle}\label{Sec:4.1}

We begin our exploration with snapshots of density fields from a gravity current evolving in an ambient pulsating flow with $Fr_o=1$, $KC_b=50$, and $Fr_m=0.4$. We examine the snapshots taken at various phases of the oscillation cycle and compare them with similar snapshots of density fields obtained from a gravity current propagating under exclusively oscillatory forcing ($Fr_o=1$, $KC_b=50$, and $Fr_m=0$), see Figure 4 of the study by \citet{Bingol2025}.  

\begin{figure}[ht!]
  \caption{\label{fig:Figure 10} Dimensionless density fields are presented for $Fr_m=0.4$, $Fr_o=1$, and $KC_b=50$ at different phases of the imposed ambient flow ($L_{AR}\approx 3.7$): panel (a) $\phi = 0^\circ$, (b) $\phi = 90^\circ$, (c) $\phi = 180^\circ$, (d) $\phi = 270^\circ$, and (e) $\phi = 360^\circ$. Red arrows depict the horizontal velocity $u(y)$ of the imposed pulsating flow profile. The value of the density (with $0 \le \rho \le 1$) is indicated by the color bar.
  }
  \hspace*{-3mm}  
  \includegraphics[width=170mm]{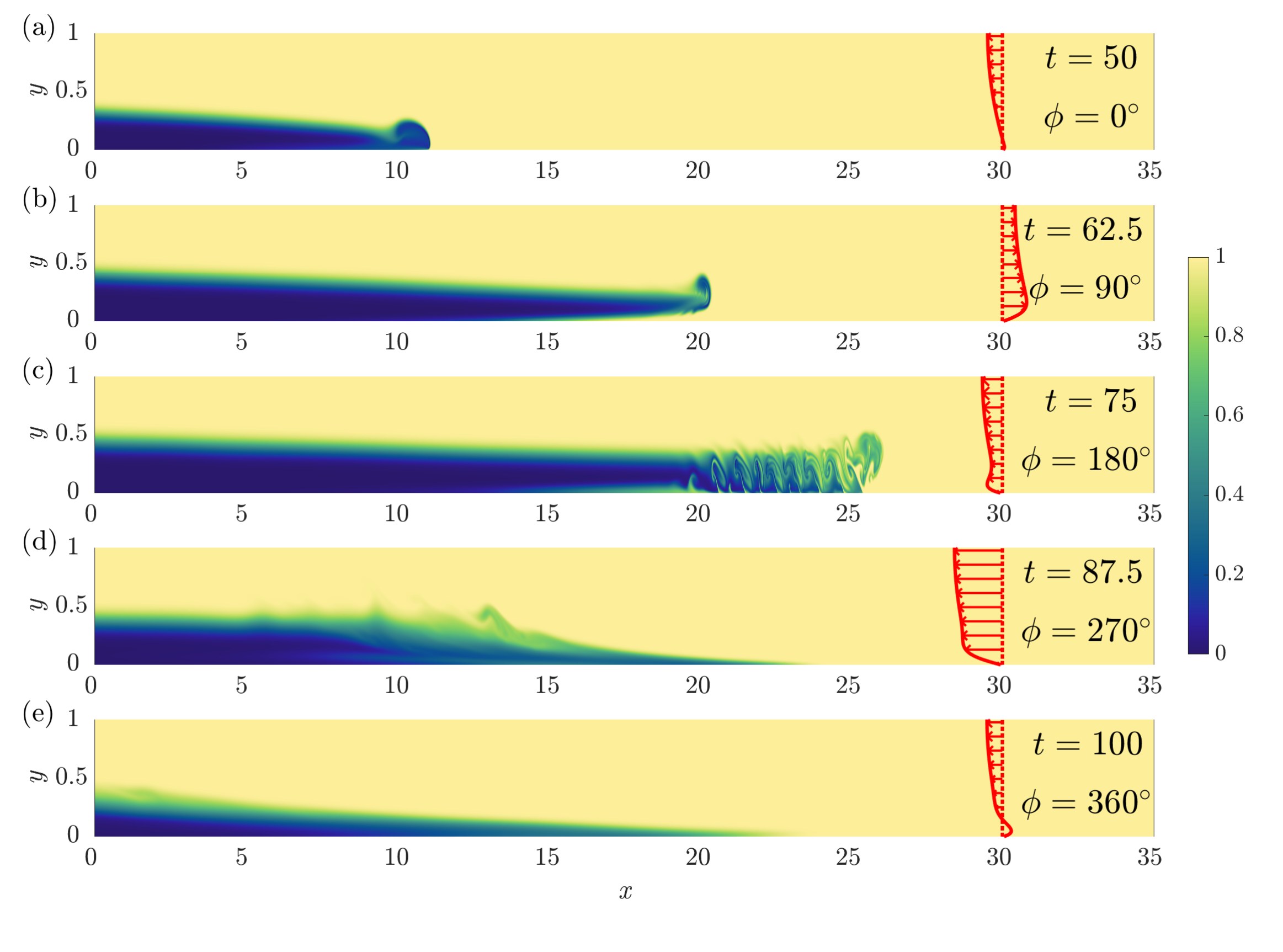}
\end{figure}

\begin{figure}[ht!]
  \caption{\label{fig:Figure 11} (a) The maximum lifting area $A_{l,max}$ calculated for each oscillation cycle for simulations with $KC_b=50$ and $Fr_o=1$ and $Fr_m\in (0, 0.1, 0.2, 0.3, 0.4)$. (b) Scaling of the average maximum lifting area $\langle A_{l,max}\rangle$ for the same set of simulations. The black dashed line indicates the scaling $\langle A_{l,max} \rangle \approx 1.3-3.1 Fr_m$. The fit is based on linear least squares from the data for $Fr_m$.}
  \hspace*{-3mm}  
  \includegraphics[width=150mm]{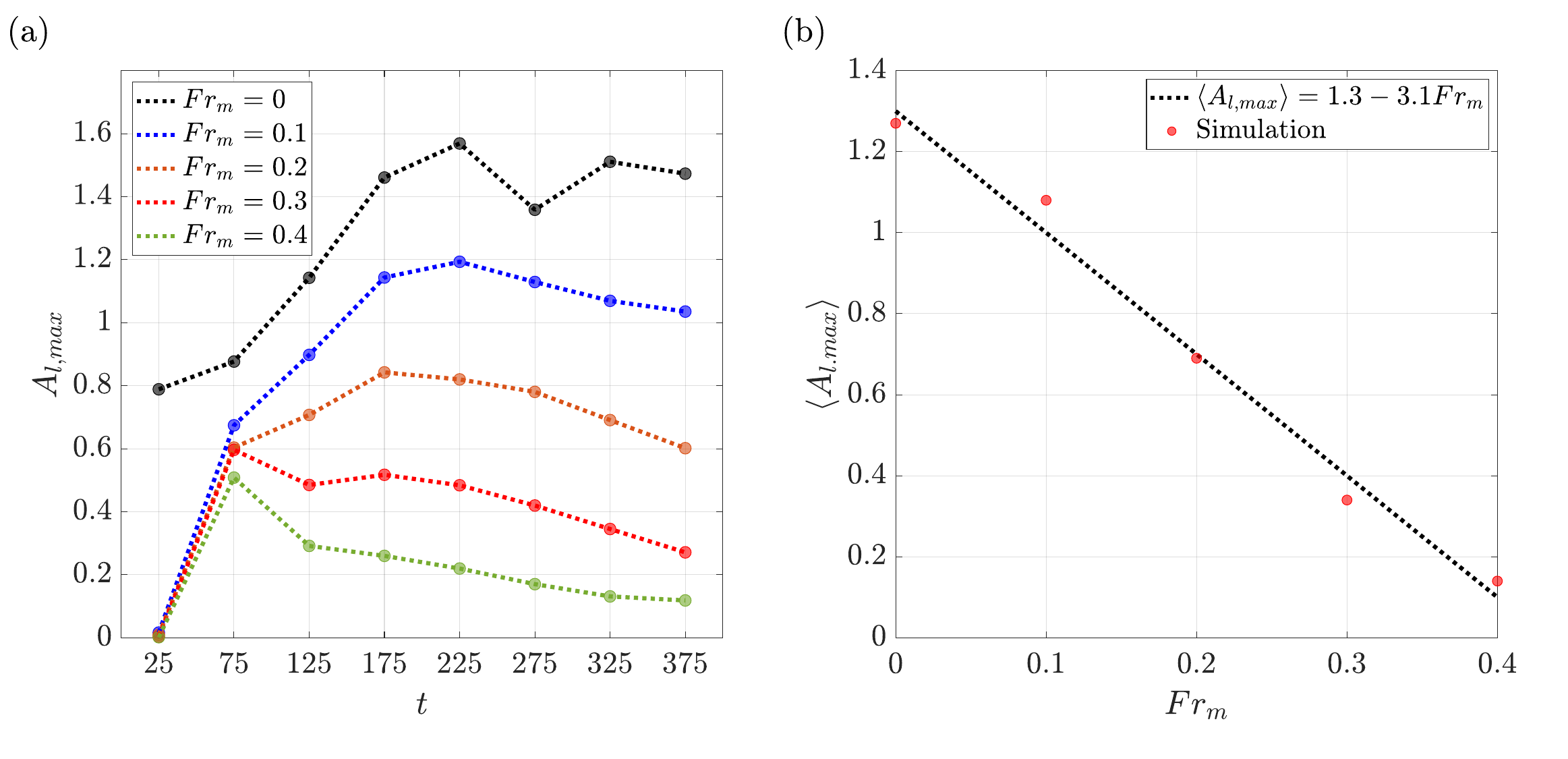}
\end{figure}

At $t=50$, we observe a distinct head of the gravity current (see figure \ref{fig:Figure 10}(a)). During the first quarter of the oscillation cycle ($0^\circ\!\le\!\phi\!\le\!90^\circ$), the direction of the external pulsating flow field is generally aligned with the propagation direction of the gravity current. The dimensionless free-stream velocity, representative of the flow velocity at the stress-free top surface, varies between $U(t)/U_b=-\frac{3}{2}Fr_m$ for $\phi=0^\circ$ and $U(t)/U_b=Fr_o-\frac{3}{2}Fr_m$ for $\phi=90^\circ$, and the phase $\phi_+$ at which the free-stream velocity changes from opposing to aligning with the propagation direction of the gravity current is $\phi_+=\sin^{-1}(\frac{3}{2}Fr_m/Fr_o)\approx 37^\circ$. The imposed velocity profile of the pulsating flow varies non-monotonically with height (see the red arrows in figure \ref{fig:Figure 10}). The height-dependent velocity profile promotes differential advection, causing the dense current to move over the lighter fluid. This process is known as lifting of the gravity current, and the area beneath the gravity current is termed the lifting area. We adopt the definition of lifting height $h_l(x,t)$ as proposed by \citet{Bingol2025}, given as

\begin{equation}
    h_l (x,t) = h_{cm}(x,t)   - 2 \, \frac{\int_{0}^{h_{cm}(x,t)} \rho (x,y,t) \, (h_{cm}(x,t)-y) \, dy}{\int_{0}^{h_{cm}(x,t)} \rho (x,y,t) \, dy}~, \label{lift-height}
\end{equation}

\noindent which is used to determine the lifting area $A_l$ by integrating the lifting height in the channel direction. It is expressed as

\begin{equation}
\label{eq:lifting_area_eq}
    A_l(t) = \int_{X_{0}}^{X_{fr}} h_l(x,t) \, dx~,
\end{equation}

\noindent with $h_l=0$ for $x\ge X_{fr}$, and $X_0=-80$ representing the left boundary of the domain. Initiation of strong density redistributions in the head of the gravity current occurs approximately at the instant when the lifting area attains its maximum value, denoted by $A_{l,max}$. The maximum lifting area occurs for every oscillation cycle and is clearly present for all $Fr_m$, as shown in figure \ref{fig:Figure 11}(a). However, it decreases almost linearly with $Fr_m$ as expected, because differential advection is reduced with increased opposing flow. The average maximum lifting area $\langle A_{l,max} \rangle$ is calculated for oscillation cycles starting at $t=200$ (to avoid transient effects; see figure \ref{fig:Figure 11}(a)). For $\langle A_{l,max}\rangle$ as function of $Fr_m$ (for $Fr_o=1$ and $KC_b=50$), including the case without mean opposing flow ($Fr_m=0$) from \citet{Bingol2025}, we obtain $\langle A_{l,max} \rangle \approx 1.3-(3.1\pm 0.2) Fr_m$; see figure \ref{fig:Figure 11}(b). This relation suggests that lifting will disappear for $Fr_m\gtrsim 0.42$. 

In the second quarter of the oscillation cycle ($90^\circ\!\le\!\phi\!\le\!180^\circ$), the oscillatory flow component remains aligned with the propagation direction of the gravity current. However, during this quarter, at $\phi_-=\sin^{-1}(\frac{3}{2}Fr_m/Fr_o)\approx 143^\circ$, the pulsating flow reverses direction again (from aligning with to opposing the propagation of the gravity current) and does not drive differential advection. The unstable situation of dense fluid above light fluid results in a significant (vertical) redistribution of high- and low-density fluid (figure \ref{fig:Figure 10}(c)) and suggests the presence of a Rayleigh–Taylor (RT)
instability. \citet{Bingol2025} observed similar density redistributions for gravity currents solely under oscillatory forcing, but at a later stage during the oscillation cycle, with a phase delay of approximately $30^\circ$ (compare figure \ref{fig:Figure 10}(c) with figure 4(c) in \citet{Bingol2025}). In the third quarter of the oscillation cycle ($180^\circ\!\le\!\phi\!\le\!270^\circ$), the density redistribution caused by the lifting of the current is completed (see figure \ref{fig:Figure 10}(d)). During the fourth quarter ($270^\circ\!\le\!\phi\!\le\!360^\circ$), the current height in the front region decreases significantly. This is due to the ambient velocity profile that acts against the propagation of the gravity current (see the red arrows in figure \ref{fig:Figure 10}(d) and (e)).

\subsection{Front position and propagation velocity under pulsating flow conditions}\label{Sec-4.2}

The propagation properties of the gravity current are characterized by measuring the position of the front $X_{fr}(t)$, the propagation velocity $U_{fr}(t)$, and the horizontal position of the normalized center of mass $L_m(t)$ of the dense fluid on the right side of the gate. We are particularly interested in the cycle-averaged versions of these quantities.

\begin{figure}[ht!]
  \caption{\label{fig:Figure 12} The cycle-averaged front position, $\langle X_{fr} \rangle_{n}$, across different cycles for varying $Fr_o$ (a), and the difference between these values to the averaged front position for $Fr_o=0$ (b). The different colors indicate $Fr_o$, and different marker distinguish $Fr_m$. Each $\langle X_{fr} \rangle_{n}$ (and similar for $\langle X_{fr} \rangle_{n}-\langle X_{fr} \rangle_{n,0}$) is plotted at time $t=(n+\frac{1}{2})KC_b$, thus halfway its averaging window, see Eq. (\ref{cycle-averaged-XF}).}
  \hspace*{-3mm}  
  \includegraphics[width=150mm]{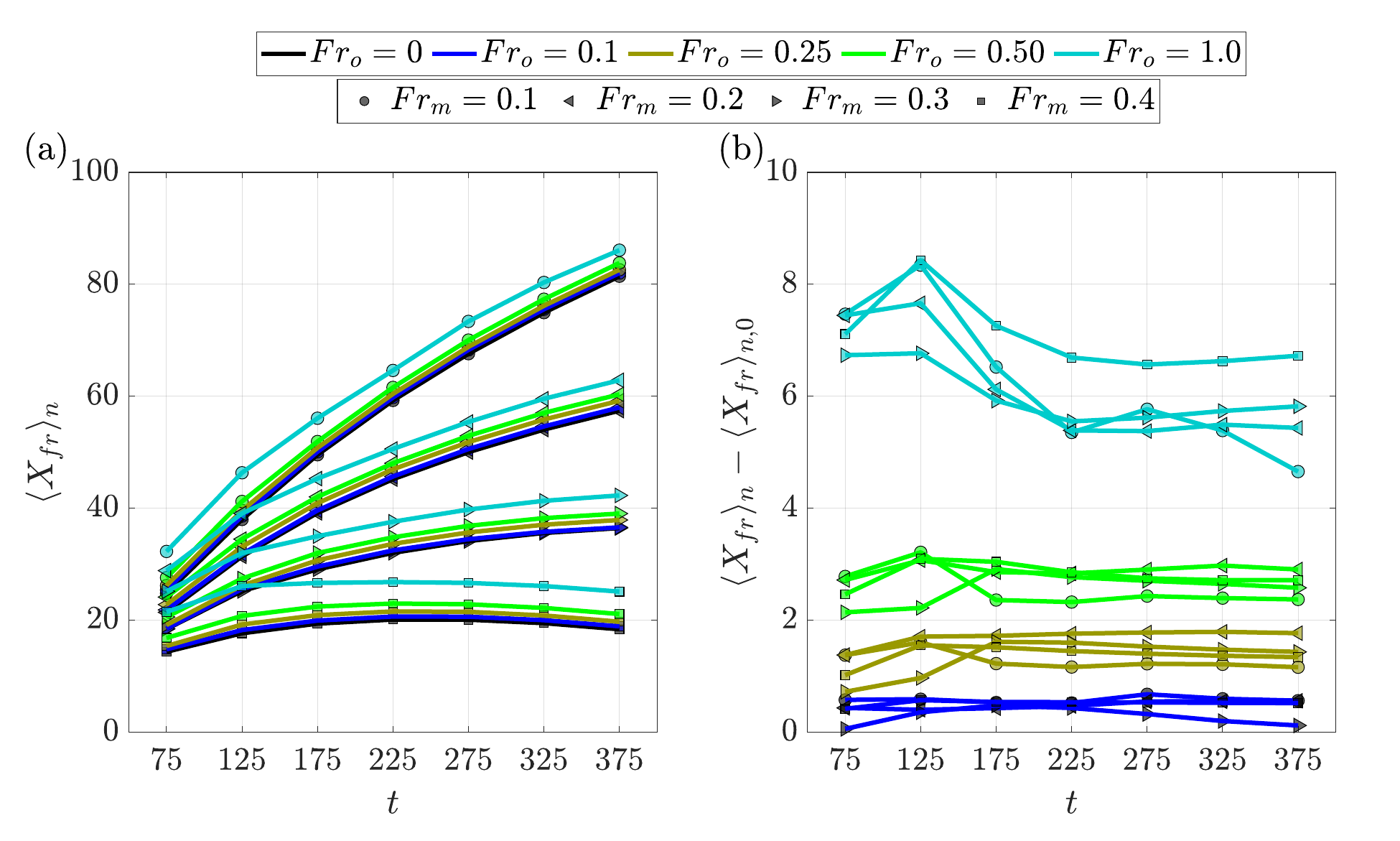}
\end{figure}

We quantify the impact of the ambient oscillatory flow component on $X_{fr}(t)$ with the following cycle-averaged front position,

\begin{equation}\label{cycle-averaged-XF}
\langle X_{fr}\rangle_n = \sqrt{\frac{1}{KC_b}\int_{nKC_b}^{(n+1)KC_b} X_{fr}^2(t)dt}~,
\end{equation}

\noindent with $1\!\le\! n\!\le\! 7$. From figure \ref{fig:Figure 12}(a), we can conclude that the cycle-averaged front position is mainly determined by the mean opposing flow, quantified by $Fr_m$. With increasing $Fr_m$, the propagation of the gravity current decreases (and becomes approximately arrested for $Fr_m=0.3$ and arrested for $Fr_m=0.4$). Increasing the amplitude of the oscillating component of the ambient flow slightly enhances $\langle X_{fr}\rangle_n$, but the general behavior remains similar, independent of $Fr_o$. This suggests that the impact of the mean and oscillating part of the ambient flow on front propagation is almost completely decoupled in the current parameter range of $(Fr_m, Fr_o)$. This is confirmed by the data presented in figure \ref{fig:Figure 12}(b) where we show the difference between the cycle-averaged front position under pulsating-flow conditions corrected for the front position with the same $Fr_m$ and $Fr_o=0$, $\langle X_{fr}\rangle_{n}-\langle X_{fr}\rangle_{n,0}$. This difference in front propagation for each $Fr_o$ is hardly affected by the magnitude of the mean flow component $Fr_m$.   

\begin{figure}[ht!]
  \caption{\label{fig:Figure 13} (a) The period-averaged front velocity, $\langle U_{fr} \rangle_n$, with $n=1, 2, \cdot \cdot, 7$, for $Fr_o$ and $Fr_m$. The different colors indicate $Fr_o$, and different markers distinguish $Fr_m$. Each $\langle U_{fr} \rangle_{n}$ is plotted at time $t=nKC_b$, see Eq. (\ref{ufrn}). (b)  Mean values of $\langle U_{fr} \rangle_n$ over cycles $4$ to $7$, denoted as $\langle\!\langle U_{fr} \rangle\!\rangle$, as function of $Fr_m$. The black dashed line indicates the scaling $\langle\!\langle U_{fr} \rangle\!\rangle \approx 0.21-0.55 Fr_m$. The fit is based on linear least squares from the data for $Fr_m$.}
  \hspace*{-3mm}  
  \includegraphics[width=150mm]{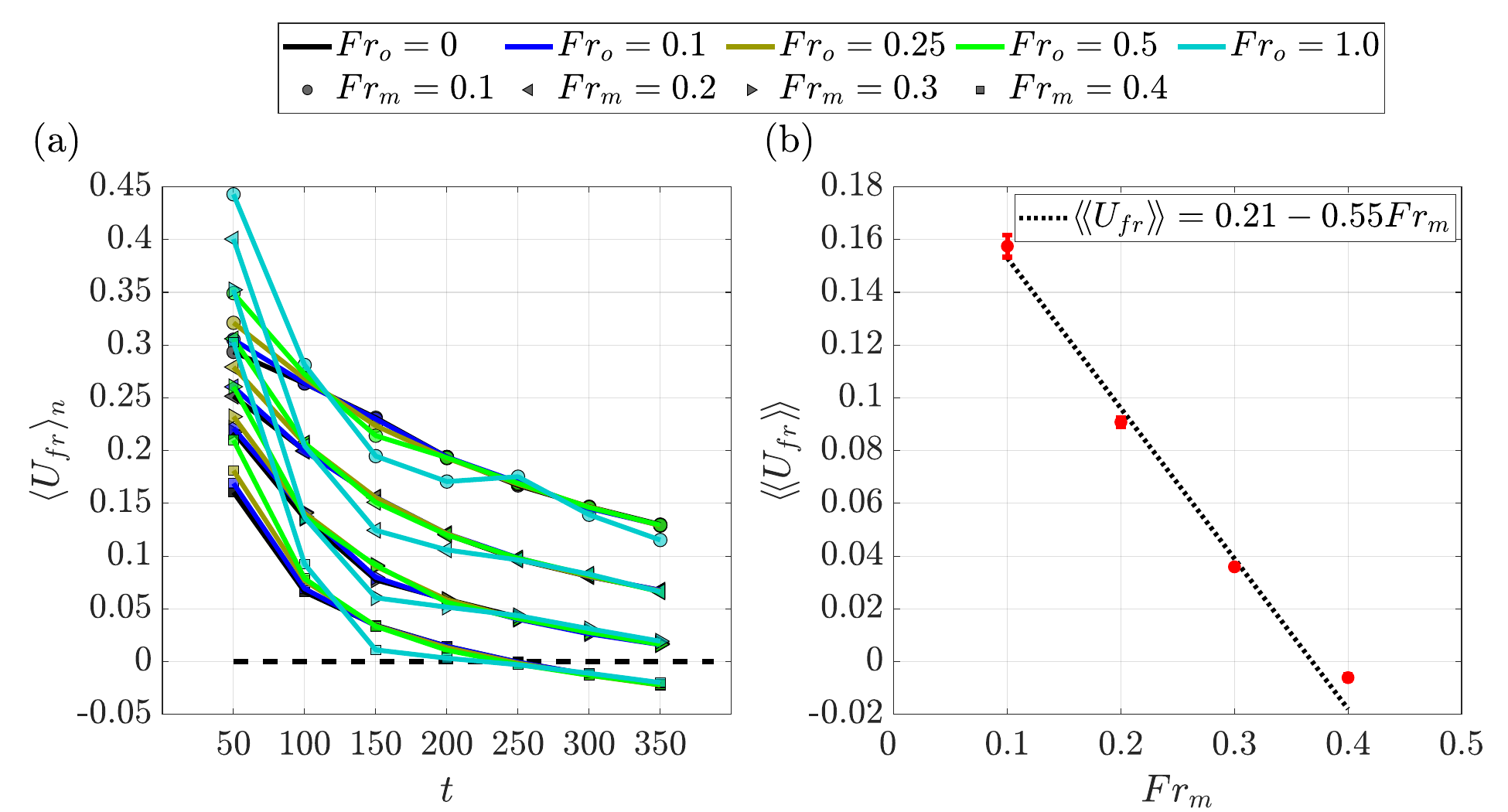}
\end{figure}

The cycle-averaged front propagation velocity is defined as
\begin{equation}
    \langle U_{fr} \rangle_n= \frac{1}{KC_b}\left(\langle X_{fr} \rangle_{n} -\langle X_{fr} \rangle_{n-1}\right)~,\label{ufrn}
\end{equation}
with $1\!\le\! n\!\le\! 7$ and $\langle X_{fr} \rangle_{0}$ the average front position in the first period without oscillatory forcing. From figure \ref{fig:Figure 13}, we see that $\langle U_{fr} \rangle_n$ remains positive throughout the duration of the simulation for $Fr_m \le 0.3$. For $Fr_m=0.4$, the gravity current is almost arrested for $200 \lesssim t \lesssim 400$. The current is even pushed back somewhat by the strong ambient mean flow for $t\ge 275$. This latter phenomenon is not entirely surprising because the front position cannot remain steady. Eventually, it must become negative, since the gravity current experiences a diminishing buoyancy-driven pressure gradient over time due to dissipative effects, while the magnitude of the mean opposing flow remains constant. The data in figure \ref{fig:Figure 13} show that $\langle U_{fr} \rangle_n$ (and $\langle\!\langle U_{fr}\rangle\!\rangle$) depends almost linearly on $Fr_m$, and that variations in $Fr_o$ have only a limited impact on the cycle-averaged front propagation for $t\gtrsim 100$. 

\begin{figure}[ht!]
  \caption{\label{fig:Figure 14} (a) The cycle-averaged center of mass normalized with the front position $\langle L_m \rangle_n$, across different cycles for $Fr_o=0.1$ (circles, solid lines) and $0.4$ (squares, dashed lines), and $Fr_m\in\{0.1, 0.2, 0.3, 0.4\}$. Each $\langle L_m \rangle_{n}$ is plotted at time $t=(n+\frac{1}{2})KC_b$, thus halfway its averaging window. (b) The averaged values of $\langle L_m\rangle_n$ for $4 \le n \le 7$, denoted as $\langle\!\langle L_m\rangle\!\rangle$.}
  \hspace*{-3mm}  
  \includegraphics[width=150mm]{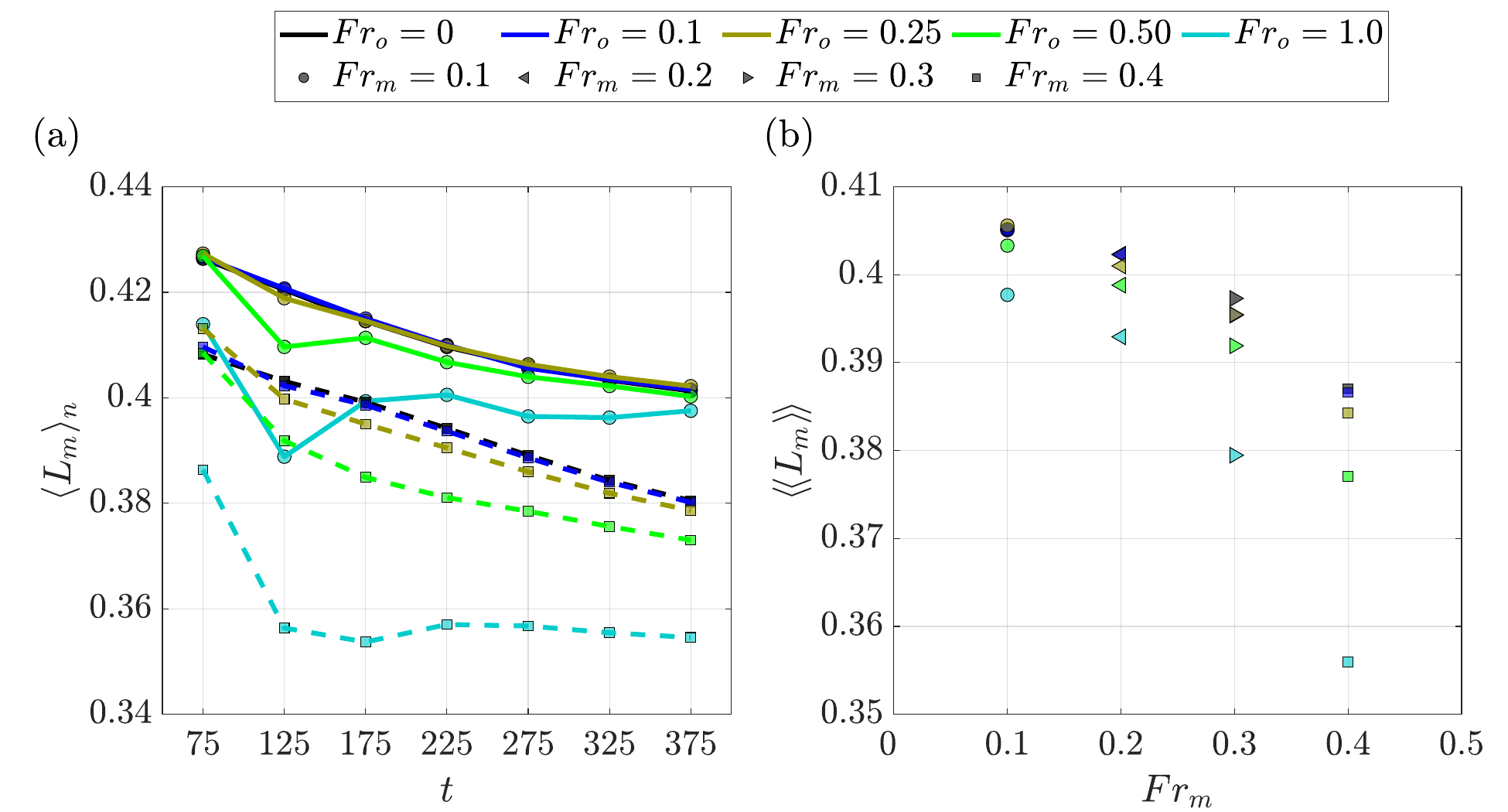}
\end{figure}

To illustrate the impact of the oscillatory component of the ambient flow on density transport, we define the cycle-averaged center of mass $\langle L_m \rangle_n$, with $L_m(t)$ defined in Eq. (\ref{equation:L_m}), in a manner similar to the cycle-averaged front position $\langle X_{fr} \rangle_n$, see Eq. (\ref{cycle-averaged-XF}). This relative measure reflects the horizontal advection of the dense fluid within the gravity current. We show $\langle L_m \rangle_n$, for $1\le n\le 7$, in figure \ref{fig:Figure 14} with $Fr_m=0.1$ and $0.4$ and $Fr_o\in\{0,0.1,0.25,0.5,1\}$. The decrease in $\langle L_m\rangle_n$ is consistent with the general observation of horizontal tailward transport of dense fluid, to be discussed in more detail in Section \ref{Sec-4.3}. Additionally, a larger $Fr_m$ or larger $Fr_o$ will increase horizontal transport towards the tail of the gravity current. For a concise comparison of $\langle L_m\rangle_n$, we calculate the average $\langle\!\langle L_m\rangle\!\rangle = [\sum_{n=4}^7\langle L_m\rangle_n]/4$ for different values of $Fr_o$ and $Fr_m$ (see figure \ref{fig:Figure 14}(b)), avoiding the strongest transient effects observed for $t\lesssim 200$. The variation caused by $Fr_o$ is comparable to that caused by $Fr_m$, indicating that the oscillatory and mean components of the pulsating flow significantly influence horizontal density transport within the gravity current.

A general conclusion is that (beyond the initial transient effect for $t\lesssim 100$) the cycle-averaged front position and propagation velocity of the gravity current depends on $Fr_m$ (the mean ambient flow), but is hardly affected by $Fr_o$, the oscillating component of the ambient flow. However, horizontal density transport is affected by both $Fr_m$ and $Fr_o$.

\subsection{Cumulative mass transport by non-hydrostatic processes}\label{Sec-4.3}

As illustrated in Section \ref{Sec:4.1}, the gravity current experiences additional density redistributions caused by the oscillatory component of the pulsating flow. The density transport that occurs along the channel relative to the front position, denoted by $\xi=x-X_{fr}(t)$, can be quantified by introducing the mass flow with respect to the front velocity, given as

\begin{equation}\label{equation:q}
q (\xi,t) = \int_{0}^{1} [u(\xi,y,t)-U_{fr}(t)] \, \rho (\xi,y,t) \, dy~.
\end{equation}

\begin{figure}[ht!]
  \caption{\label{fig:Figure 15} The total mass transport $\langle Q (\xi)\rangle$ at the front region of the gravity current (with $Fr_m=0.4$ and $Fr_o=1$) with respect to the front velocity, averaged over $1 \leq n \leq 7$ ($50 \leq t \leq 400$), is shown. The blue, red, and black curves represent the cumulative mass transport during the first half ($0^\circ\!\le\!\phi\!\le\!180^\circ$) and the second half ($180^\circ\!\le\!\phi\!\le\!360^\circ$) of the oscillation cycle, and the full cycle ($0^\circ\!\leq\!\phi\!\leq\!360^\circ$), respectively.}
\begin{tikzpicture}
\node at (0,0) {\includegraphics[width=80mm]{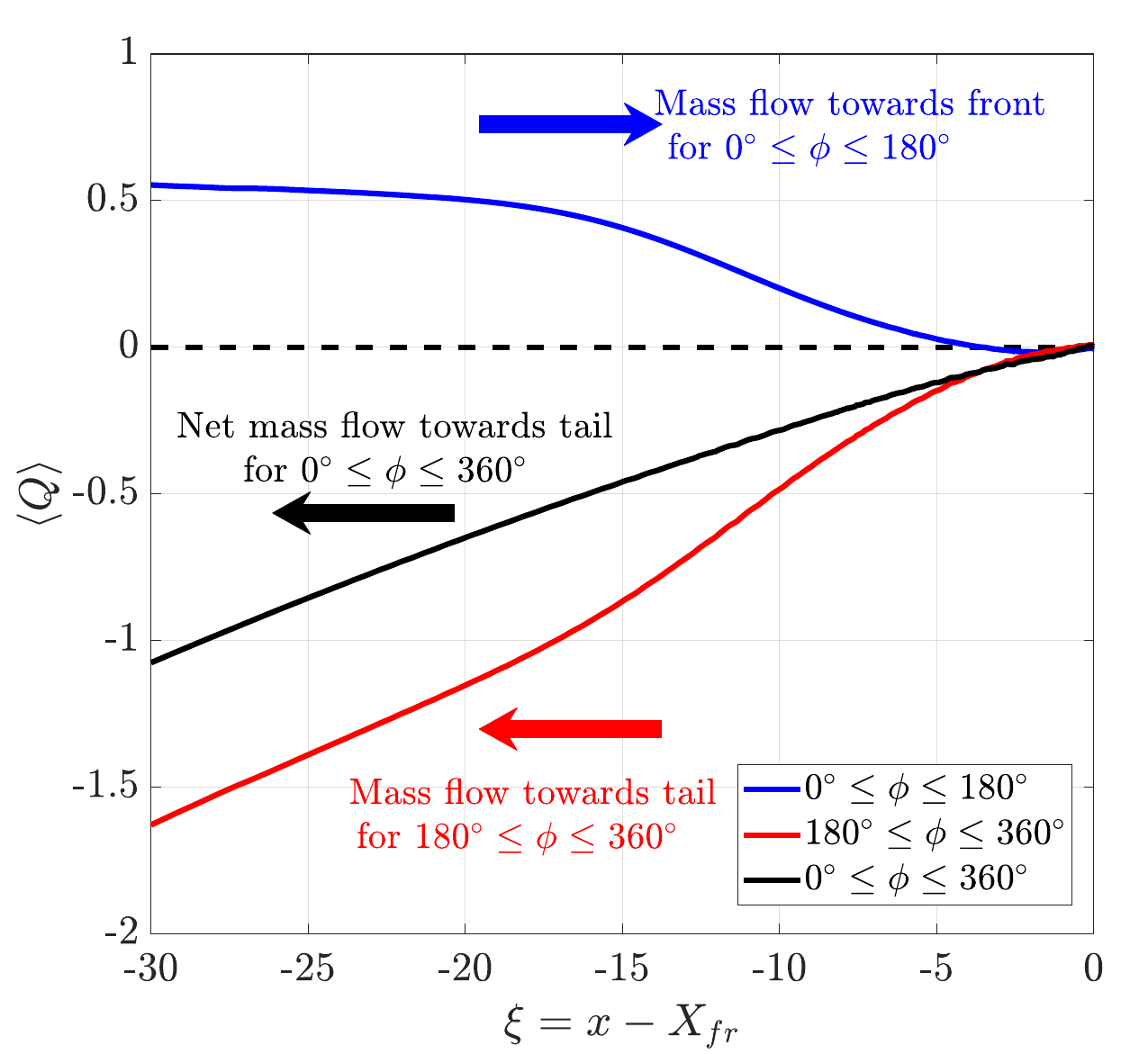}};
\end{tikzpicture}
\end{figure}

\noindent The cumulative mass transport that occurs in different phases of the oscillation cycle relative to the front position is calculated according to

\begin{equation}
Q_n (\xi) = \int_{(n+p_1)KC_b}^{(n+p_2)KC_b} q (\xi,t) dt ~,\label{eq:Qn}
\end{equation}

\noindent where $p_1=0$ and $p_2=1/2$ for the first half of the oscillation cycle ($0^\circ\!\le\!\phi\!\le\!180^\circ$); $p_1=1/2$ and $p_2=1$ for the second half of the oscillation cycle ($180^\circ\!\le\!\phi\!\le\!360^\circ$); and $p_1=0$ and $p_2=1$ for the entire oscillation cycle ($0^\circ\!\le\!\phi\!\le\!360^\circ$). We calculate the cumulative mass transport averaged for all cycles within the simulation duration and denote this quantity by $\langle Q(\xi)\rangle = [\sum_{n=1}^7Q_n(\xi)]/7$. To illustrate cumulative mass transport, we first present the results for one specific case ($Fr_m=0.4$ and $Fr_o=1$) in figure \ref{fig:Figure 15} using the three versions of Eq. (\ref{eq:Qn}). During the first half of the oscillation cycle ($0^\circ\!\le\!\phi\!\le\!180^\circ$) the cumulative mass transport (blue curve) is directed towards the front of the gravity current. In contrast, during the second half ($180^\circ\!\le\!\phi\!\le\!360^\circ$), the cumulative mass transport (red curve) reverses toward the tail, driven by the externally imposed pulsating velocity profile with the mean and oscillating ambient flow opposing gravity current propagation. The cycle-averaged mass transport over the full oscillation cycle ($0^\circ\!\le\!\phi\!\le\!360^\circ$) reveals net mass transport towards the tail of the gravity current (see the black lines in figure \ref{fig:Figure 15}).

\begin{table}[!ht]
\caption{\label{table:suppl-movies}
Overview of the parameter settings ($Fr_m$, $Fr_o$) used in the supplementary movies discussed in Section \ref{Sec-4.3}.}
\begin{ruledtabular}
\begin{tabular}{lcc}
Supplementary movie & $Fr_m$ & $Fr_o$ \\
\hline
Movie 6  & 0 & 0 \\
Movie 7  & 0 & 0.5 \\
Movie 8  & 0.1 & 0.5 \\
Movie 9  & 0.1 & 1.0 \\
Movie 10 & 0.2 & 0.5 \\
Movie 11 & 0.2 & 1.0 \\
Movie 12 & 0.3 & 1.0 \\
\end{tabular}
\end{ruledtabular}
\end{table}

\begin{figure}[ht!]
  \caption{\label{fig:Figure 16} Averaged cumulative mass transport with respect to the front velocity, $\langle Q(\xi)\rangle$, averaged over the seven oscillation cycles between $t=50$ and $400$, for the first half (blue curves), second half (red curves) and entire (black curves) oscillation cycle. For $Fr_o=0$ the same averaging windows are used as for the pulsating cases. In all panels we show $\langle Q(\xi)\rangle$ for $Fr_o=0$ (dotted curves), $Fr_o=0.5$ (dashed curves) and $Fr_o=1$ (solid curves). (a) $Fr_m=0.1$, (b) $Fr_m=0.2$, (c) $Fr_m=0.3$ and (d) $Fr_m=0.4$.}
\begin{tikzpicture}
\node at (0,0) {\includegraphics[width=110mm]{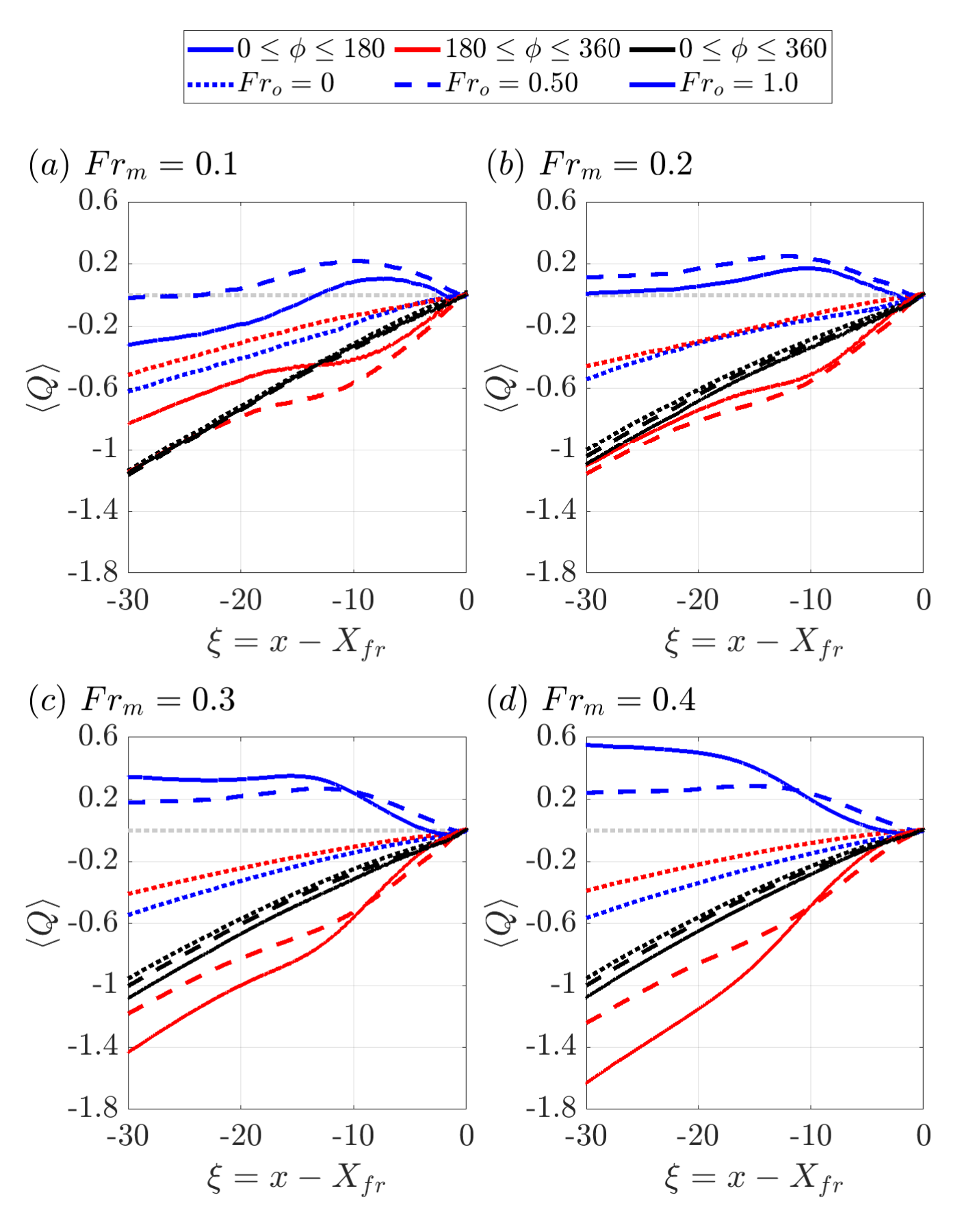}};
\end{tikzpicture}
\end{figure}

Cumulative mass transport has been explored for all combinations with $Fr_o\in\{0, 0.25, 0.5, 1\}$ and $Fr_m\in\{0.1, 0.2, 0.3, 0.4\}$. We have added seven supplementary movies for illustration of the evolution of the gravity current and associated mass transport, see Table \ref{table:suppl-movies}. An overview of the results of the cumulative mass transport for a few representative cases is provided in figure \ref{fig:Figure 16}. A first general and remarkable observation is the following: the cumulative mass transport over the entire oscillation cycle hardly depends on $Fr_m$, and larger $Fr_o$ results at best in somewhat larger net density transport in the tailward direction; see the black curves in figure \ref{fig:Figure 16}(a)-(d). However, there are clear distinctions for the cumulative mass transport for different $Fr_m$ and $Fr_o$ computed for the first and second half of the oscillation cycles. This hints at substantial redistribution of mass in the individual half-cycles, see the solid and dashed blue lines in figure \ref{fig:Figure 16}.

For $Fr_m\ge 0.3$, the cumulative mass transport during the first half, the second half, and the total cycle is similar, but overall a somewhat decreased cumulative mass transport for $Fr_o=0.5$ compared to $Fr_o=1.0$. For $0^\circ\!\le\!\phi\!\le\!180^\circ$, $\langle Q(\xi)\rangle$ remains positive. However, for $Fr_m\lesssim 0.2$, the cumulative mass transport during the first half of the oscillation cycle is strongly reduced and eventually becomes negative compared to the propagation of the front; see figure \ref{fig:Figure 16}(a) and \ref{fig:Figure 16}(b). This indicates that, even when the ambient flow aligns with the propagation direction of the gravity current, there is tailward density transport within the gravity current. For $Fr_m=0.2$, this is the case for $17^\circ\!\lesssim\! \phi\!\lesssim\! 163^\circ$, compared to $37^\circ\!\lesssim\! \phi\!\lesssim\! 143^\circ$ for $Fr_m=0.4$. Positive cumulative mass transport towards the front of the gravity current only survives near the gravity current head. For $Fr_o=0$ the cumulative mass transport compared to the front position is always negative; see the dotted curves in figure \ref{fig:Figure 16}(a)-(d). The ambient flow never aligns with the propagation direction of the gravity current. 

These observations with regard to the cumulative mass transport for $0^\circ\!\le\!\phi\!\le\!180^\circ$ and $180^\circ\!\le\!\phi\!\le\!360^\circ$, correlate nicely with the presence or absence of KH billows, see Section \ref{Sect-3.1}, and Rayleigh-Taylor-like (RT) instabilities, respectively. We first consider the cases $Fr_m=0$, with $Fr_o=0$ and $0.5$ (see the supplementary movies). In both cases, we observe the formation of KH billows that all have a clear tendency to move away from the gravity current head. This implies coherent advective transport of high-density fluid patches in the negative $x$ direction. This appears to be the basic state, and for the case with $Fr_o=0.5$, the alignment of the ambient flow with the propagation direction of the gravity current during the first half of the oscillation cycle does not counteract this drift of the KH billows sufficiently well. For $Fr_o=0$, the drift of the KH billows, present only for $Fr_m\lesssim 0.2$, is directed away from the gravity current front. This drift is enhanced due to the ambient mean flow in the negative $x$ direction (Section \ref{Sect-3Gen}) and $\langle Q\rangle$ is always negative. The KH billows generated in the simulations with $Fr_m=0.1$, and $Fr_o=0.5$ and $Fr_o=1.0$, respectively, have a clear tendency to drift in the negative $x$ direction, with respect to the front position $X_{fr}(t)$. This explains the observation that $\langle Q\rangle\lesssim 0$ for $0^\circ\!\lesssim\!\phi\!\lesssim\!180^\circ$ in figure \ref{fig:Figure 16}(a) (blue curves). The cases with $Fr_m=0.2$, and $Fr_o=0.5$ and $1.0$, respectively, show at best a tendency of the KH billows to drift away from the gravity current head. Additionally, the number of KH billows generated is significantly reduced. This explains why for these cases $\langle Q\rangle\gtrsim 0$ for $0^\circ\!\lesssim\!\phi\!\lesssim\!180^\circ$, see figure \ref{fig:Figure 16}(b) (blue curves): less KH billows for coherent transport of high-density fluid in the negative $x$ direction. However, when KH instabilities and KH billow formation are suppressed, for $Fr_m\gtrsim 0.3$, heavy fluid is contained in the core of the gravity current and tends to be more easily transported horizontally in the current toward the front of the gravity current due to the aligned ambient flow during the first half of the oscillation cycle. 

For $Fr_m=0$ and $Fr_o=0.5$, we observe another interesting phenomenon: due to the presence of lifting, a result of differential advection, a thin layer of light fluid is generated below the gravity current head (see supplementary movie 7), a configuration that is intrinsically unstable. When the ambient oscillating flow changes direction (at $\phi=180^\circ$) this configuration becomes unstable and RT-like instability occurs with strong vertical mixing of light and heavy fluid. Similar behavior can be found for $Fr_m\ge 0.1$ and $Fr_o=1$ (strong instability similar to RT that induces vertical density redistribution) and for $Fr_m\ge 0.1$ and $Fr_o=0.5$ (relatively weak instability similar to RT and associated vertical density redistributions). This instability and associated density redistributions typically occur for $120^\circ\!\lesssim\!\phi\!\lesssim\!240^\circ$, see the supplementary movies 8-12.

The presence of KH and RT-like instabilities has a distinct effect on mixing efficiency. Following the approach used by \citet{Peltier2003,Salehipour2016} and \citet{Agrawal2021} we can introduce the ratio ${\mathcal{M}}/D_p$ of the irreversible mixing rate ${\mathcal{M}}$ and the laminar diffusion rate $D_p$ (the 2D versions of the expressions introduced by \citet{Salehipour2016}). The generation of KH billows initially results in a gradual increase of the normalized irreversible mixing rate ${\mathcal{M}}/D_p$ over time (and its value is a qualitative measure of the abundance and strength of KH billows). Later, the ratio decreases as the KH billows slowly disappear. For example, the cases with opposing mean flow and $Fr_o=0$ give the following values: ${\mathcal{M}}/D_p\approx 4$ for $Fr_m=0$, ${\mathcal{M}}/D_p\approx 3$ for $Fr_m=0.1$, and ${\mathcal{M}}/D_p\approx 2.5$ for $Fr_m=0.2$, consistent with the reduction of the prominence of KH billows with increasing $Fr_m$, see Section \ref{Sect-3.1}. The cases with pulsating ambient flow clearly show the impact of RT-like instabilities on the normalized irreversible mixing rate ${\mathcal{M}}/D_p$, in particular for the cases with $Fr_o=0.5$ and $1.0$. The baseline is comparable with the cases discussed for opposing mean flow and $Fr_o=0$, and reflects the presence of KH billows and associated mixing processes. However, episodes around $t=75$, $120$ and $160$ (all coinciding with periods of strong RT-like instabilities) are observed to significantly enhance ${\mathcal{M}}/D_p$ values. For example, for $Fr_m=0.1$, we observe peak values ${\mathcal{M}}/D_p\approx 5$ for $Fr_o=1.0$ and ${\mathcal{M}}/D_p\approx 4$ for $Fr_o=0.5$ at $t\approx 75$. For $Fr_m=0.3$, we find ${\mathcal{M}}/D_p\approx 4$ at $t\approx 75$ and a slightly smaller value at $t\approx 120$, both for $Fr_o=1.0$, while for $Fr_o=0$, we have ${\mathcal{M}}/D_p\approx 2$. The normalized irreversible mixing rate clearly distinguishes between mixing induced by KH billows and during episodes with RT-like instabilities, the latter resulting in significantly stronger density redistributions.

Finally, the presence or absence of KH billows and RT-like instabilities near the bottom in the gravity current head, or in other words the presence or absence of strong non-hydrostatic effects, with associated vertical density transport, has significant implications for horizontal density transport over large distances. 

\section{Summary and conclusions} \label{sec:summary}

We investigated how pulsating flow conditions influence various aspects of gravity currents, including front position and propagation velocity of the gravity current, gravity current thickness, lifting of the gravity current, horizontal advective density transport by Kelvin-Helmholtz billows, and strong and violent vertical density redistributions in the gravity current head due to Rayleigh-Taylor-like instabilities. Our study considered both individual and combined effects of the mean and oscillating components of the pulsating flow. Phenomena such as lifting, the emergence of KH billows and their dynamics, and the presence of RT-like instabilities are supported by 12 supplementary movies. To clearly delineate their separate and combined effects on gravity currents, we naturally divided our study into two parts, focusing on mean and pulsating flow conditions. 

In the first part of our study, we focused on gravity currents evolving in an opposing mean flow, without an oscillating component. 
For mean opposing flows, with $Fr_m\le 0.2$ shear-driven KH instabilities emerge at the interface of heavy and light fluid near the gravity current head, generating KH billows that slowly drift against the propagation direction of the gravity current, advecting patches with heavy fluid towards the tail of the gravity current. They thus play a crucial role in horizontal density transport. For $Fr_m\ge 0.3$ no KH billows are formed. The presence or absence of KH instabilities is supported by a qualitative analysis based on the local gradient Richardson number. The front position and propagation velocity of the gravity currents satisfy the classical scaling regimes: acceleration, slumping, inertial, and viscous phases \citep{Cantero2007OnCurrents}, and the mean ambient flow affects both the thickness of the gravity current and the timing of the transitions between the phases. The emergence of the KH billows coincides with the inertial phase, and their generation ceases when the gravity current enters the viscous phase. The absence of the inertial phase for $Fr_m=0.4$ perfectly aligns with the absence of KH instabilities. Finally, the amount of heavy fluid flowing through the gate, normalized with the front position, decreases with increasing $Fr_m$.

In the second part of our study, we analyzed the gravity current dynamics under ambient pulsating-flow conditions. Our qualitative observations of the density field evolution indicate that density redistributions under oscillating and pulsating ambient flow share many similarities, such as generating KH billows and lifting of the gravity current head. For pulsating flow conditions with $Fr_m\le 0.2$, we observe the emergence of KH billows and the associated advective transport of coherent patches with heavy fluid in a direction opposite to the propagation of the gravity current. The formation of these KH billows is less regular compared to the case with $Fr_o=0$ and absent for $Fr_m\ge 0.3$. Lifting of the gravity current head provides a second instability mechanism, because as a result of differential advection a layer of light fluid remains stuck below the heavy fluid contained by the gravity current head \citep{Bingol2025}. As long as the gravity current accelerates over the layer with light fluid, the configuration remains stable, but when it decelerates (when the oscillating component of the pulsating ambient flow changes from being aligned to being opposed to the direction of the gravity current propagation), Rayleigh-Taylor-like instabilities emerge. When this occurs violent vertical redistribution of heavy and light fluid occurs, as can be observed in the supplementary movies, and contributes to horizontal density transport (because of the presence of KH billows for $Fr_m\lesssim 0.2$ and mean ambient flow in the negative $x$ direction). Our analysis also shows that the position of the front and the velocity of the gravity currents are minimally affected by the oscillatory component of the pulsating flow. However, the mass flow in the along-channel direction indicates a net density transport from the front towards the tail of the current, with a noticeable increase as oscillatory forcing strengthens, which enhances RT-like instabilities in the gravity current head. 

In summary, our study highlights the significant effects of pulsating flow on gravity currents, with distinct influences from both the mean and oscillating components. The mean opposing flow reduces front velocity, current height, and the generation of instabilities at the interface, leading to an earlier shift to a viscous propagation phase. In contrast, the oscillatory component of the ambient pulsating flow has minimal influence on the front position and velocity of the gravity current for time scales larger than the oscillation period. The mixing process is accelerated for gravity currents with pulsating flow compared to that with oscillating flow, and there is enhanced mass transport toward the tail of the current compared to only mean flow. More generally, we can conclude that non-hydrostatic effects, such as the presence of KH billows and the role of RT-like instabilities, with associated vertical density transport, have significant implications for horizontal density transport over large distances. These phenomena should be included in the modeling of salt intrusions in rivers and estuaries.

In the present study, we have chosen a fixed Reynolds number, $Re=3000$, to allow a ($Fr_o,Fr_m$) parameter study with available computational resources and to facilitate comparisons with previous work reported by \citet{Bingol2025}. To assess the sensitivity of our results to $Re$, we performed three simulations with $Re=6000$ and $Sc=5$ (using a grid size $\delta_x=\delta_y=0.005$). Two of them with only opposing mean flow ($Fr_m=0.2$ and $0.4$) and one with $Fr_m=0.4$ and $Fr_o=1.0$. All the trends observed in these higher Reynolds number simulations are similar to those for the same parameter settings and $Re=3000$. The most significant difference is that inertial effects impact the dynamics of gravity currents with $Re=6000$, and associated formation of KH billows, for a much longer duration. For example, for $Fr_m=0.2$ and $Re=6000$, it is observed that the formation of KH billows is absent for $t\gtrsim 265$, while for the same $Fr_m$ but $Re=3000$, the formation of KH billows disappears for $t\gtrsim 85$. In agreement with these observations, the viscous regime is established at later times for all three cases with $Re=6000$ and the front velocity $U_{fr}(t)$ is higher for $Re=6000$. Finally, lifting is somewhat more significant for $Re=6000$, with $Fr_m=0.4$ and $Fr_o=1.0$, and the normalized irreversible mixing rate ${\mathcal{M}}/D_p$ over time shows a similar behavior for $Re=6000$ compared to $Re=3000$, including a clear signature of the role of RT-like instabilities, for $Fr_m=0.4$ and $Fr_o=1.0$. We can conclude that the dynamical processes (including the emergence of different propagation phases) during the propagation of the gravity current against constant and pulsating counter flows, the position and velocity of the front, the associated mass transport and lifting phenomena are not highly sensitive for the Reynolds number in the range $3000\lesssim Re\lesssim 6000$.

A natural next step beyond the present 2D DNS is the extension to fully 3D simulations, which are essential for capturing inherently 3D instabilities such as lobe-and-cleft structures, spanwise variations of lifting, and natural breakdown of Kelvin-Helmholtz billows. Although 2D models provide valuable information on the combined effects of mean and pulsating counter flows, 3D simulations can provide further information on the fine-scale (lateral) density redistribution, enhanced entrainment, and mixing processes driven by fully 3D turbulence. Extending this framework to 3D DNS at higher Reynolds numbers can allow for an exploration of the role of vorticity in shaping mixing and lifting, and the spanwise characteristics of lobe-and-cleft instabilities. These directions should form the focus of future work on this topic.
\vspace{0.5cm}

\noindent \textbf{Declaration of interest:} The authors report no conflict of interest.
\vspace{2mm}

\noindent \textbf{Author ORCIDs:} 

\noindent Cem Bingol https://orcid.org/0000-0002-8436-0470

\noindent Matias Duran Matute https://orcid.org/0000-0002-1340-339X

\noindent Eckart Meiburg https://orcid.org/0000-0003-3670-8193

\noindent Herman J. H. Clercx https://orcid.org/0000-0001-8769-0435

\begin{acknowledgments}
\noindent Eckart Meiburg gratefully acknowledges support by NSF grant NSF-1936358. The authors thank Rob Uittenbogaard for helpful discussions and suggestions.

\noindent \textbf{Funding:} This work is part of the Perspectief Program SALTIsolutions, which is financed by NWO Domain Applied and Engineering Sciences (2022/TTW/01344701) in collaboration with private and public partners. This work utilized the Dutch national e-infrastructure, supported by the SURF Cooperative through grant number PRJS-1006. \vspace{0.01mm}
\end{acknowledgments}

\bibliography{references}

@article{Biegert2017a,
    title = {{A collision model for grain-resolving simulations of flows over dense, mobile, polydisperse granular sediment beds}},
    year = {2017},
    journal = {Journal of Computational Physics},
    author = {Biegert, E. and Vowinckel, B. and Meiburg, E.},
    pages = {105--127},
    volume = {340},
    issn = {10902716}
}

@article{deNijs2011,
    title = {{Advection of the salt wedge and evolution of the internal flow structure in the rotterdam waterway}},
    year = {2011},
    journal = {Journal of Physical Oceanography},
    author = {de Nijs, M. A.J. and Pietrzak, J. D. and Winterwerp, J. C.},
    number = {1},
    volume = {41},
    issn = {00223670}
}

@article{Hartel2000,
    title = {{Analysis and direct numerical simulation of the flow at a gravity-current head. Part 1. Flow topology and front speed for slip and no-slip boundaries}},
    year = {2000},
    journal = {Journal of Fluid Mechanics},
    author = {H{\"{a}}rtel, C. and Meiburg, E. and Necker, F.},
    pages = {189--212},
    volume = {418}
}

@article{Moin1985,
    title = {{Application of a fractional-step method to incompressible Navier-Stokes equations}},
    year = {1985},
    journal = {Journal of Computational Physics},
    author = {Kim, J. and Moin, P.},
    number = {2},
    pages = {308--323},
    volume = {59},
    issn = {10902716}
}

@article{Dai2021,
    title = {{Boussinesq and non-Boussinesq gravity currents propagating on unbounded uniform slopes in the deceleration phase}},
    year = {2021},
    journal = {Journal of Fluid Mechanics},
    author = {Dai, Albert and Huang, Yu Lin},
    volume = {917},
    issn = {14697645}
}

@article{Stancanelli2018,
    title = {{Computational fluid dynamics for modeling gravity currents in the presence of oscillatory ambient flow}},
    year = {2018},
    journal = {Water (Switzerland)},
    author = {Stancanelli, L. M. and Musumeci, R. E. and Foti, E.},
    number = {5},
    volume = {10},
    issn = {20734441}
}

@article{Riddel1970,
    title = {{Densimetric exchange flow in rectnagular channels IV - The arrested saline wedge}},
    year = {1970},
    journal = {La Houille Blanche},
    author = {Riddell, J. F.},
    number = {4},
    month = {6},
    pages = {317--330},
    volume = {56},
    publisher = {Informa UK Limited},
    issn = {0018-6368}
}

@article{Palomar2010,
    title = {{Desalination in Spain: Recent developments and recommendations}},
    year = {2010},
    journal = {Desalination},
    author = {Palomar, P. and Losada, I. J.},
    number = {1-3},
    volume = {255},
    issn = {00119164}
}

@article{Martin2019,
    title = {{Development of gravity currents on slopes under different interfacial instability conditions}},
    year = {2019},
    journal = {Journal of Fluid Mechanics},
    author = {Martin, Antoine and Negretti, M. Eletta and Hopfinger, E. J.},
    volume = {880},
    issn = {14697645}
}

@article{Maggi2025b,
    title = {{Dynamics and mixing of gravity currents over an array of cylindrical obstacles}},
    year = {2025},
    journal = {Physics of Fluids},
    author = {Maggi, M. R. and Di Lollo, G. and Adduce, C.},
    number = {7},
    volume = {37},
    publisher = {AIP Publishing LLC},
    url = {https://doi.org/10.1063/5.0276373},
    issn = {10897666}
}

@article{Craske2015JFM,
    title = {{Energy dispersion in turbulent jets. Part 1. Direct simulation of steady and unsteady jets}},
    year = {2015},
    journal = {Journal of Fluid Mechanics},
    author = {Craske, J. and Van Reeuwijk, M.},
    pages = {500--537},
    volume = {763},
    issn = {14697645}
}

@article{Blanchette2006,
    title = {{Evaluation of a simplified approach for simulating gravity currents over slopes of varying angles}},
    year = {2006},
    journal = {Computers and Fluids},
    author = {Blanchette, F. and Piche, V. and Meiburg, E. and Strauss, M.},
    number = {5},
    pages = {492--500},
    volume = {35},
    isbn = {7737025863},
    issn = {00457930}
}

@article{Dai2020,
    title = {{Experiments on gravity currents propagating on unbounded uniform slopes}},
    year = {2020},
    journal = {Environmental Fluid Mechanics},
    author = {Dai, Albert and Huang, Yu Lin},
    number = {6},
    volume = {20},
    issn = {15731510}
}

@book{Ungarish2020,
    title = {{Gravity currents and intrusions: Analysis and prediction}},
    year = {2020},
    booktitle = {Gravity Currents And Intrusions: Analysis And Prediction},
    author = {Ungarish, M.}
}

@article{Kollner2020b,
    title = {{Gravity currents over fixed beds of monodisperse spheres}},
    year = {2020},
    journal = {Journal of Fluid Mechanics},
    author = {K{\"{o}}llner, T. and Meredith, A. and Nokes, R. and Meiburg, E.},
    volume = {901},
    issn = {14697645}
}

@article{Shin2004,
    title = {{Gravity currents produced by lock exchange}},
    year = {2004},
    journal = {Journal of Fluid Mechanics},
    author = {Shin, J. O. and Dalziel, S. B. and Linden, P. F.},
    pages = {1--34},
    volume = {521},
    issn = {00221120}
}

@article{Tokyay2011b,
    title = {{Gravity currents propagating over periodic arrays of blunt obstacles: Effect of the obstacle size}},
    year = {2011},
    journal = {Journal of Fluids and Structures},
    author = {Tokyay, T. and Constantinescu, G. and Gonzalez-Juez, E. and Meiburg, E.},
    number = {5-6},
    pages = {798--806},
    volume = {27},
    issn = {08899746}
}

@article{Bingol2025,
    title = {{Gravity currents under oscillatory forcing}},
    year = {2025},
    journal = {Journal of Fluid Mechanics},
    author = {Bingol, C. and Duran-Matute, M. and Zhu, R. and Meiburg, E. and Clercx, H. J.H.},
    month = {12},
    volume = {1002},
    publisher = {Cambridge University Press},
    issn = {14697645}
}

@misc{Huppert2006,
    title = {{Gravity currents: A personal perspective}},
    year = {2006},
    booktitle = {Journal of Fluid Mechanics},
    author = {Huppert, H. E.},
    volume = {554},
    issn = {14697645}
}

@article{Harten1997,
    title = {{High resolution schemes for hyperbolic conservation laws}},
    year = {1997},
    journal = {Journal of Computational Physics},
    author = {Harten, A.},
    number = {2},
    pages = {260--278},
    volume = {135},
    issn = {00219991}
}

@article{Necker2002,
    title = {{High-resolution simulations of particle-driven gravity currents}},
    year = {2002},
    journal = {International Journal of Multiphase Flow},
    author = {Necker, F. and H{\"{a}}rtel, C. and Kleiser, L. and Meiburg, E.},
    pages = {279--300},
    volume = {28},
    issn = {18063691}
}

@article{Biegert2017b,
    title = {{High-resolution simulations of turbidity currents}},
    year = {2017},
    journal = {Progress in Earth and Planetary Science},
    author = {Biegert, E. and Vowinckel, B. and Ouillon, R. and Meiburg, E.},
    number = {1},
    volume = {4},
    issn = {21974284}
}

@article{Birman2007a,
    title = {{Lock-exchange flows in sloping channels}},
    year = {2007},
    journal = {Journal of Fluid Mechanics},
    author = {Birman, V. K. and Battandier, B. A. and Meiburg, E. and Linden, P. F.},
    pages = {53--77},
    volume = {577},
    isbn = {0022112006},
    issn = {00221120}
}

@article{Maggi2022,
    title = {{Lock-release gravity currents propagating over roughness elements}},
    year = {2022},
    journal = {Environmental Fluid Mechanics},
    author = {Maggi, Maria Rita and Adduce, Claudia and Negretti, Maria Eletta},
    number = {2-3},
    pages = {383--402},
    volume = {22},
    publisher = {Springer Netherlands},
    url = {https://doi.org/10.1007/s10652-022-09845-6},
    isbn = {0123456789},
    issn = {15731510}
}

@article{Long1956,
    title = {{Long Waves In A Two-Fluid System}},
    year = {1956},
    journal = {Journal of Meteorology},
    author = {Long, R. R.},
    number = {1},
    volume = {13},
    issn = {0095-9634}
}

@article{Necker2005,
    title = {{Mixing and dissipation in particle-driven gravity currents}},
    year = {2005},
    journal = {Journal of Fluid Mechanics},
    author = {Necker, F. and H{\"{a}}rtel, C. and Kleiser, L. and Meiburg, E.},
    pages = {339--372},
    volume = {545},
    issn = {00221120}
}

@article{VanReeuwijk2019JFM,
    title = {{Mixing and entrainment are suppressed in inclined gravity currents}},
    year = {2019},
    journal = {Journal of Fluid Mechanics},
    author = {Van Reeuwijk, M. and Holzner, M. and Caulfield, C. P.},
    pages = {786--815},
    volume = {873},
    issn = {14697645},
    arxivId = {1808.08980}
}

@article{Peltier2003,
    title = {{Mixing efficiency in stratified shear flows}},
    year = {2003},
    journal = {Annual Review of Fluid Mechanics},
    author = {Peltier, W. R. and Caulfield, C. P.},
    volume = {35},
    issn = {00664189}
}

@article{Meiburg2015,
    title = {{Modeling gravity and turbidity currents: Computational approaches and challenges}},
    year = {2015},
    journal = {Applied Mechanics Reviews},
    author = {Meiburg, E. and Radhakrishnan, S. and Nasr-Azadani, M.},
    number = {4},
    pages = {1--23},
    volume = {67},
    issn = {00036900}
}

@article{Zemach2019,
    title = {{On gravity currents of fixed volume that encounter a down-slope or up-slope bottom}},
    year = {2019},
    journal = {Physics of Fluids},
    author = {Zemach, T. and Ungarish, M. and Martin, A. and Negretti, M. E.},
    number = {9},
    volume = {31},
    issn = {10897666}
}

@article{Kokkinos2023_JHR,
    title = {{On the dynamics of gravity current motion in a stratified ambient}},
    year = {2023},
    journal = {Journal of Hydraulic Research},
    author = {Kokkinos, A. and Prinos, P.},
    number = {5},
    pages = {703--719},
    volume = {61},
    publisher = {Taylor and Francis Ltd.},
    issn = {00221686}
}

@article{Zahtila2024_PoF,
    title = {{On the propagation of planar gravity currents into a stratified ambient}},
    year = {2024},
    journal = {Physics of Fluids},
    author = {Zahtila, T. and Lam, W. K. and Chan, L. and Sutherland, D. and Moinuddin, K. and Dai, A. and Skvortsov, A. and Manasseh, R. and Ooi, Andrew},
    number = {3},
    month = {3},
    volume = {36},
    publisher = {American Institute of Physics},
    issn = {10897666}
}

@article{Agrawal2021,
    title = {{Probing the high mixing efficiency events in a lock-exchange flow through simultaneous velocity and temperature measurements}},
    year = {2021},
    journal = {Physics of Fluids},
    author = {Agrawal, Tanmay and Ramesh, Bhaarath and Zimmerman, Spencer J. and Philip, Jimmy and Klewicki, Joseph C.},
    number = {1},
    volume = {33},
    issn = {10897666}
}

@article{Martin2020,
    title = {{Propagation of a continuously supplied gravity current head down bottom slopes}},
    year = {2020},
    journal = {Physical Review Fluids},
    author = {Martin, A. and Negretti, M. E. and Ungarish, M. and Zemach, T.},
    number = {5},
    volume = {5},
    issn = {2469990X}
}

@incollection{Geyer2011,
    title = {{The Dynamics of Strongly Stratified Estuaries}},
    year = {2011},
    booktitle = {Treatise on Estuarine and Coastal Science},
    author = {Geyer, W. R. and Ralston, D. K.},
    month = {3},
    pages = {37--51},
    volume = {2},
    publisher = {Elsevier Inc.},
    isbn = {9780080878850}
}

@article{Anjum2013,
    title = {{The instantaneous Froude number and depth of unsteady gravity currents}},
    year = {2013},
    journal = {Journal of Hydraulic Research},
    author = {Anjum, H. J. and McElwaine, J. N. and Caulfield, C. C. P.},
    number = {4},
    pages = {432--445},
    volume = {51},
    issn = {00221686}
}

@article{Maxworthy2002a,
    title = {{The propagation of a gravity current into a linearly stratified fluid}},
    year = {2002},
    journal = {Journal of Fluid Mechanics},
    author = {Maxworthy, T. and Leilich, J. and Simpson, J. E. and Meiburg, E. H.},
    number = {May},
    pages = {371--394},
    volume = {453},
    issn = {00221120}
}

@article{NasrAzadani2011,
    title = {{TURBINS: An immersed boundary, Navier-Stokes code for the simulation of gravity and turbidity currents interacting with complex topographies}},
    year = {2011},
    journal = {Computers and Fluids},
    author = {Nasr-Azadani, M. M. and Meiburg, E.},
    number = {1},
    pages = {14--28},
    volume = {45},
    issn = {00457930}
}

@article{Maggi2023b,
    title = {{Turbulence characteristics and mixing properties of gravity currents over complex topography}},
    year = {2023},
    journal = {Physics of Fluids},
    author = {Maggi, Maria Rita and Negretti, M. Eletta and Hopfinger, Emil J. and Adduce, Claudia},
    number = {1},
    volume = {35},
    issn = {10897666}
}

@article{Ralston2017,
    title = {{Turbulent and numerical mixing in a salt wedge estuary: Dependence on grid resolution, bottom roughness, and turbulence closure}},
    year = {2017},
    journal = {Journal of Geophysical Research: Oceans},
    author = {Ralston, D. K. and Cowles, G. W. and Geyer, W. R. and Holleman, R. C.},
    number = {1},
    month = {1},
    pages = {692--712},
    volume = {122},
    publisher = {Blackwell Publishing Ltd},
    issn = {21699291}
}

@article{Salehipour2016,
    title = {{Turbulent mixing due to the Holmboe wave instability at high Reynolds number}},
    year = {2016},
    journal = {Journal of Fluid Mechanics},
    author = {Salehipour, Hesam and Caulfield, C. P. and Peltier, W. R.},
    volume = {803},
    issn = {14697645}
}

@article{Hallez2009ACurrents,
    title = {{A numerical investigation of horizontal viscous gravity currents}},
    year = {2009},
    journal = {Journal of Fluid Mechanics},
    author = {Hallez, Y. and Magnaudet, J.},
    volume = {630},
    issn = {14697645}
}

@article{Stancanelli2017ACoast,
    title = {{A small scale Pressure Retarded Osmosis power plant: dynamics of the brackish effluent discharge along the coast}},
    year = {2017},
    journal = {Ocean Engineering},
    author = {Stancanelli, L. M. and Musumeci, R. E. and Cavallaro, L. and Foti, E.},
    volume = {130},
    issn = {00298018}
}

@article{Kaptein2019EffectRegime,
    title = {{Effect of the water depth on oscillatory flows over a flat plate: from the intermittent towards the fully turbulent regime}},
    year = {2019},
    journal = {Environmental Fluid Mechanics},
    author = {Kaptein, S. J. and Duran-Matute, M. and Roman, F. and Armenio, V. and Clercx, H. J.H.},
    number = {5},
    volume = {19},
    issn = {15731510}
}

@article{Hallworth1996EntrainmentCurrents,
    title = {{Entrainment into two-dimensional and axisymmetric turbulent gravity currents}},
    year = {1996},
    journal = {Journal of Fluid Mechanics},
    author = {Hallworth, M. A. and Huppert, H. E. and Phillips, J. C. and Sparks, R. S. J.},
    pages = {289--311},
    volume = {308},
    issn = {00221120}
}

@article{Britter1978ExperimentsHead,
    title = {{Experiments on the dynamics of a gravity current head}},
    year = {1978},
    journal = {Journal of Fluid Mechanics},
    author = {Britter, R. E. and Simpson, J. E.},
    number = {2},
    pages = {223--240},
    volume = {88},
    issn = {14697645}
}

@article{Ouillon2021GravitySources,
    title = {{Gravity currents from moving sources}},
    year = {2021},
    journal = {Journal of Fluid Mechanics},
    author = {Ouillon, R. and Kakoutas, C. and Meiburg, E. and Peacock, T.},
    volume = {924},
    issn = {14697645}
}

@article{Martinsen1987ImplementationModel,
    title = {{Implementation and testing of a lateral boundary scheme as an open boundary condition in a barotropic ocean model}},
    year = {1987},
    journal = {Coastal Engineering},
    author = {Martinsen, E. A. and Engedahl, H.},
    number = {5-6},
    volume = {11},
    issn = {03783839}
}

@book{Saad2003IterativeSystems,
    title = {{Iterative Methods for Sparse Linear Systems}},
    year = {2003},
    author = {Saad, Y.},
    edition = {Second}
}

@article{Gayen2010LargeCurrent,
    title = {{Large eddy simulation of a stratified boundary layer under an oscillatory current}},
    year = {2010},
    journal = {Journal of Fluid Mechanics},
    author = {Gayen, B. and Sarkar, S. and Taylor, J.H.},
    pages = {233--266},
    volume = {643}
}

@article{Smyth2000LengthLayers,
    title = {{Length scales of turbulence in stably stratified mixing layers}},
    year = {2000},
    journal = {Physics of Fluids},
    author = {Smyth, W. D. and Moum, J. N.},
    number = {6},
    volume = {12},
    issn = {10706631}
}

@article{Chorin1968NumericalEquations,
    title = {{Numerical Solution of the Navier-Stokes Equations}},
    year = {1968},
    journal = {Mathematics of Computation},
    author = {Chorin, A. J.},
    number = {104},
    volume = {22},
    issn = {00255718}
}

@article{Hoult1972OilSea,
    title = {{Oil Spreading on the Sea}},
    year = {1972},
    journal = {Annual Review of Fluid Mechanics},
    author = {Hoult, D. P.},
    number = {1},
    volume = {4},
    issn = {0066-4189}
}

@article{Cantero2007OnCurrents,
    title = {{On the front velocity of gravity currents}},
    year = {2007},
    journal = {Journal of Fluid Mechanics},
    author = {Cantero, M. I. and Lee, J. R. and Balachandar, S. and Garcia, M. H.},
    volume = {586},
    issn = {14697645}
}

@article{Miles1961OnFlows,
    title = {{On the stability of heterogeneous shear flows}},
    year = {1961},
    journal = {Journal of Fluid Mechanics},
    author = {Miles, J. W.},
    number = {4},
    volume = {10},
    issn = {14697645}
}

@article{Zhu2021RemovalCurrent,
    title = {{Removal of a dense bottom layer by a gravity current}},
    year = {2021},
    journal = {Journal of Fluid Mechanics},
    author = {Zhu, R. and He, Z. and Meiburg, E.},
    volume = {916},
    issn = {14697645}
}

@article{Mashayek2017RoleLayers,
    title = {{Role of overturns in optimal mixing in stratified mixing layers}},
    year = {2017},
    journal = {Journal of Fluid Mechanics},
    author = {Mashayek, A. and Caulfield, C. P. and Peltier, W. R.},
    volume = {826},
    issn = {14697645}
}

@article{Atoufi2023StratifiedInstabilities,
    title = {{Stratified inclined duct: Two-layer hydraulics and instabilities}},
    year = {2023},
    journal = {Journal of Fluid Mechanics},
    author = {Atoufi, A. and Zhu, L. and Lefauve, A. and Taylor, J. R. and Kerswell, R. R. and Dalziel, S. B. and Lawrence, G. A. and Linden, P. F.},
    volume = {977},
    issn = {14697645}
}

@article{Huppert1982TheSurface,
    title = {{The propagation of two-dimensional and axisymmetric viscous gravity currents over a rigid horizontal surface}},
    year = {1982},
    journal = {Journal of Fluid Mechanics},
    author = {Huppert, H. E.},
    volume = {121},
    issn = {14697645}
}

@article{Lawrence1991TheInterface,
    title = {{The stability of a sheared density interface}},
    year = {1991},
    journal = {Physics of Fluids A},
    author = {Lawrence, G. A. and Browand, F. K. and Redekopp, L. G.},
    number = {10},
    volume = {3},
    issn = {08998213}
}

@article{Ouillon2019TurbidityFluid,
    title = {{Turbidity currents propagating down a slope into a stratified saline ambient fluid}},
    year = {2019},
    journal = {Environmental Fluid Mechanics},
    author = {Ouillon, R. and Meiburg, E. and Sutherland, B. R.},
    number = {5},
    volume = {19},
    issn = {15731510}
}

\end{document}